\documentclass[11pt,a4paper]{article}
\usepackage{jheppub,tcolorbox}
\usepackage[T1]{fontenc}
\usepackage[utf8]{inputenc}

\newcommand{\ord}[1]{{\scriptscriptstyle(#1)}}
\newcommand{\ind}[2]{{}^{#1}_{#2}}

\title{\boldmath{Quantum Off-Shell Recursion Relation}}

 \author{Kanghoon Lee}
\affiliation{Asia Pacific Center for Theoretical Physics, Postech,\\ Pohang 37673, Korea}
\affiliation{Department of Physics, Postech, \\Pohang 37673, Korea}

\emailAdd{kanghoon.lee1@gmail.com}

\abstract{
We construct off-shell recursion relations for arbitrary loop-level scattering amplitudes beyond the conventional tree-level recursion relations for $\phi^{4}$-theory and the Yang--Mills theory. We define a quantum perturbiner expansion that includes loop corrections from the quantum effective action formalism by identifying the external source. Our method clearly shows how the perturbiner expansion becomes an off-shell current generating function. Instead of using the classical equations of motion in the conventional perturbiner method, we exploit the Dyson--Schwinger equation to derive the quantum off-shell recursion relation to arbitrary order of loop-level scattering amplitudes. We solve the recursion relation and reproduce the results which agree up to one-loop six-point scattering amplitudes for $\phi^{4}$-theory. Furthermore, we construct the recursions for computing loop-level correlation functions by replacing the choice of the external source.}

\begin{document}
\maketitle	
\flushbottom
%
\section{Introduction}
Recursion relations are a powerful tool for computing scattering amplitudes efficiently. There are two types of recursion relations: on-shell and off-shell. The on-shell recursion, so-called BCFW recursion relation \cite{Britto:2004ap,Britto:2005fq,Arkani-Hamed:2008bsc}, provides an elegant procedure for constructing the higher-point amplitudes from lower-point amplitudes by ``recycling''. It uses the analytic properties of tree-level scattering amplitudes via complex deformations of the external momenta without Feynman diagrams. On the other hand, the off-shell recursion relation by Berends and Giele is based on the recursive structure of the interaction vertices of a theory \cite{Berends:1987me}. The key ingredient is the off-shell current, representing an amplitude with an off-shell leg. 

There is an alternative method for constructing the off-shell recursion relation, the so-called perturbiner method by Rosly and Selivanov \cite{Rosly:1996vr,Rosly:1997ap,Selivanov:1997aq,Selivanov:1997an,Selivanov:1997ts}. The perturbiner method generates the same off-shell recursion relation from the perturbiner expansion, a generating function of the off-shell currents, using the classical equations of motion (EoM). Thus it provides a manifest relation between the solutions of the classical EoM and the tree-level scattering amplitudes. Compared with the derivation by Berends and Giele, the perturbiner method enables a systematic derivation of the recursion relation. This method has been successfully applied to some non-gravitational effective field theories \cite{Lee:2015upy,Mafra:2015vca,Mafra:2016mcc,Mafra:2016ltu,Mafra:2015gia,Mizera:2018jbh} and gravity \cite{Cheung:2016say,Cheung:2017kzx,Gomez:2021shh,Cheung:2021zvb,Cho:2021nim}. These recursion relations work perfectly at tree level. There have been several works to generalise the recursion relation to loop-orders \cite{Mahlon:1993fe,Mahlon:1993si,Kim:1996nd,Bern:2005cq,Arvanitakis:2019ald,Jurco:2019yfd}. 

This paper addresses a generalisation of the conventional tree-level perturbiner method to loop-level. This issue is tied to the question of how to derive the perturbiner expansion. We show that the quantum effective action formalism leads the \emph{quantum perturbiner expansion} by specifying the external source required to reproduce the LSZ reduction formula from the generating functional connected diagram. Recently, this formalism also used in the context of the off-shell recursion relation in curved spacetime \cite{Cheung:2022pdk,Herderschee:2022ntr,Armstrong:2020woi,Cohen:2022uuw}. We treat the classical field, which is defined by the vacuum expectation value of a given quantum field in the presence of the external source, as the ``field'' that we apply the perturbiner expansion. 

The conventional tree-level perturbiner method uses the classical EoM for deriving the off-shell recursion relation. Here we exploit the Dyson--Schwinger equation that is quantum analogues of the classical EoM. However, the DS equation involves functional derivatives of the fields, making it difficult to solve. Our strategy is to consider the functional derivatives as independent fields. We define them as \emph{descendant fields} and introduce their perturbiner expansion that encodes loop integrals explicitly. We generate the quantum off-shell recursion relation by substituting the quantum perturbiner expansion into the DS equation. Further, the initial condition of the recursion relation arises from the source term in the DS equation. We construct the quantum recursions for $\phi^{4}$-theory and pure Yang--Mills theory and solve the recursion for the scalar theory to one-loop level. This reproduces the same scattering amplitudes as computed using Feynman diagrams.

This formalism is not limited to scattering amplitude. We extend the quantum off-shell recursion relation for computing correlation functions. As before, we derive the perturbiner expansion for the scattering amplitudes from the quantum effective action formalism, this time to use with a different external source -- we detach the inverse propagator because we do not need an amputation. In this case, the form of the recursion relation is identical to the scattering amplitude case. However, only the initial condition is changed. We check that it gives the same correlation functions using the usual Feynman diagram computation.

The structure of this paper is as follows. In section \ref{Sec:2}, we review the quantum effective action formalism for $\phi^{4}$-theory and establish our notation. We define the quantum perturbiner expansion by specifying the external source. In Section \ref{Sec:3}, we construct the quantum off-shell recursion relation by substituting the quantum perturbiner expansion into the DS equation up to two-loop level. We derive the initial condition from the external source in the DS equation. We solve the one-loop recursion relation up to six-point amplitudes. In Section \ref{Sec:4}, we extend the recursion relation to the case of correlation functions. In section 5, we construct the quantum off-shell recursion relation for pure YM theory.

\section{Quantum Perturbiner Expansion}\label{Sec:2}
In this section, we construct the quantum perturbiner expansion for $\phi^{4}$-theory, which generalises the conventional tree-level perturbiner expansion to loop-level within the quantum effective action formalism. First, we identify the ``field'', to which we apply the quantum perturbiner expansion, with the vacuum expectation value  (VEV) of the scalar with the external source. Next, we choose the external source in a specific form to reproduce the LSZ formula from the connected generating functional. Finally, using the relation between the classical field and the connected correlation functions and the external source, we define the quantum perturbiner expansion, which provides a generating function of the quantum off-shell current. 

\subsection{Quantum effective action for $\phi^{4}$-theory}
We start by reviewing the quantum effective action for $\phi^{4}$-theory, which plays a central role in deriving the quantum perturbiner method. The classical action for $\phi^{4}$-theory with a external source $j(x)$ is given by
\begin{equation}
\begin{aligned}
  S[\phi,j] &= \int \mathrm{d}^{4}x \bigg[ -\frac{1}{2}\big(\partial_{\mu} \phi(x)\big)^{2}-\frac{1}{2} m^{2} \phi(x)^{2} - \frac{\lambda}{4!} \phi(x)^{4} + j(x) \phi(x) \bigg] \,, 
\end{aligned}\label{}
\end{equation}
where $m$ and $\lambda$ are the bare mass and the bare coupling. Throughout this paper, we denote position and momentum space integrations as follows:
\begin{equation}
  \int_{x,y\cdots} = \int \mathrm{d}^{4}x \,\mathrm{d}^{4}y \cdots 
  \quad \text{and} \quad 
  \int_{p,q,\cdots} =\int \frac{\mathrm{d}^{4}p}{(2\pi)^{4}}\frac{\mathrm{d}^{4}q}{(2\pi)^{4}}\cdots \,.
\label{}\end{equation}
We may rewrite the action in terms of the kinetic operator $K_{x,y} := K(x,y)$ and the four-point interaction vertex $V_{xyzw} := V(x,y,z,w)$,
\begin{equation}
  S[\phi,j] =  -\frac{1}{2} \int_{x,y} \phi(x) K_{x,y} \phi(y) 
  + \frac{\lambda}{4!} \int_{x,y,z,w} V_{xyzw} \phi(x) \phi(y) \phi(z) \phi(w)  
  + \int_{x}j(x) \phi(x)\,,
\label{}\end{equation}
where
\begin{equation}
\begin{aligned}
  K_{x,y} &= \big(-\Box_{y}+m^{2}\big)\delta^{4}(x-y)\,,
  \\
  V_{x,y,z,w} &= - \hat{\delta}(x-y)\hat{\delta}(x-z)\hat{\delta}(x-w) \,.
\end{aligned}\label{}
\end{equation}
The classical equation of motion (EoM) for $\phi(x)$ is given by
\begin{equation}
  \frac{\delta S[\phi,j]}{\delta \phi(x)}= -\int_{y} K(x,y) \phi(y) - \frac{\lambda}{3!} \phi(x)^{3} + j(x)=0 \,.
\label{phi4_EoM}\end{equation}

The free propagator $D_{xy}:=D(x,y)$ is given by the inverse of the kinetic operator $K_{xy}$,
\begin{equation} 
  \int_{y}K(x,y)D_{yz} = \delta^{4}(x-z)\,, 
  \qquad  
  D_{xy}=\int \frac{d^{4} p}{(2 \pi)^{4}} \tilde{D}_{p} e^{i p\cdot(x-y)} \,,
\label{scalar_propagator}\end{equation}
where $\tilde{D}_{p}$ is the free propagator in momentum space,
\begin{equation}
  \tilde{D}_{p} = \frac{1}{p^{2}+m^{2}-i \epsilon} \,.
\label{}\end{equation}
From now on, we will omit the $i\epsilon$ factor in the Feynman propagator for convenience. The dressed propagator $\mathbf{D}_{xy}$ including all-loop corrections is related to the exact two-point function,
\begin{equation} 
  \mathbf{D}_{xy} = \int_{p} e^{ip\cdot (x-y)} \tilde{\mathbf{D}}(p^{2}) = \frac{i}{\hbar}\left\langle 0|T \phi(x) \phi(y) |0\right\rangle\,.
\label{}\end{equation}
Here $\tilde{\mathbf{D}}(p^{2})$ is the momentum space dressed propagator with the self-energy $\Pi(p^{2})$,
\begin{equation}
  \tilde{\mathbf{D}}(p^{2}) = \frac{1}{p^{2}+m^{2} - \Pi(p^{2}) }\,.
\label{dressed_propagator_momentum}\end{equation}
Recall that $\Pi(p^{2})$ is given by the 1PI diagrams in the two-point function and receives loop-corrections, $\Pi = \Pi^{\ord{1}} + \Pi^{\ord{2}} +\cdots$, where 
\begin{equation}
  \frac{i}{\hbar}\Pi^{\ord{1}} = -\frac{i}{\hbar}\frac{\lambda}{2} \int_{q} \frac{-i\hbar}{q^{2}+m^{2}} = - \frac{\lambda}{2} \int_{q} \frac{1}{q^{2}+m^{2}} \,,
\label{}\end{equation}
and 
\begin{equation}
\begin{aligned}
  \frac{i}{\hbar} \Pi^{\ord{2}}(p^{2}) = \frac{\hbar}{i}\lambda^{2} \int_{q,r} 
  	\bigg(&
  	\frac{1}{4} \frac{1}{q^{2}+m^{2}} \frac{1}{(r^{2}+m^{2})^{2}}
  	\\&
  	+ \frac{1}{6} \frac{\lambda}{(q+r-p)^{2}+m^{2}} \frac{1}{q^{2}+m^{2}} \frac{1}{r^{2}+m^{2}}
  	\bigg)\,.
\end{aligned}\label{}
\end{equation}

We now introduce the generating functional for connected diagrams, $W[j]$, defined by the functional integration while keeping the external source $j(x)$
\begin{equation}
  e^{\frac{i}{\hbar}W[j]}= Z[j]= \int \mathcal{D} \phi \, e^{\frac{i}{\hbar} S[\phi,j]}\,.
\label{}\end{equation}
We may expand $W[j]$ with respect to $j_{x} := j(x)$ around $j_{x} = 0$,	
\begin{equation}
\begin{aligned}
  W[j] &= \sum_{n=2}^{\infty} \frac{1}{n!} \int_{x_{1}x_{2}\cdots x_{n}} 
  	\frac{\delta^{n}W[j]}{\delta j_{x_{1}} \delta j_{x_{2}} \cdots \delta j_{x_{n}}}\bigg|_{j=0}j_{x_{1}} j_{x_{2}}\cdots j_{x_{n}}
  	\\
  	&= -i\hbar \sum_{n=2}^{\infty} \frac{1}{n!} \int_{x_{1}x_{2}\cdots x_{n}} 
  	G_{c}(x_{1},x_{2},\cdots,x_{n}) \frac{i j_{x_{1}}}{\hbar} 	\frac{i j_{x_{2}}}{\hbar} \cdots \frac{i j_{x_{n}}}{\hbar} \,,
\end{aligned}\label{WJ_scalar}
\end{equation}
where the coefficients of the expansion $G_{c}(x_{1},x_{2},\cdots,x_{n})$ are the connected $n$-point correlation functions by definition.

The classical field $\varphi_{j}(x)$ is the VEV of the scalar field $\phi(x)$ in the presence of the external source and represented by a functional derivative of $W[j]$
\begin{equation}
\begin{aligned}
  \varphi_{x} &:= \frac{\delta W[j]}{\delta j_x} = \frac{\left\langle0| \phi(x)|0\right\rangle_{j}}{\left\langle 0|0\right\rangle_{j}}\,,
\end{aligned}\label{}
\end{equation}
where $\varphi_{x}$ and $j_{x}$ are shorthand notations for $\varphi_{j}(x)$ and $j(x)$, respectively. Using the expansion of $W[j]$ \eqref{WJ_scalar}, we may also expand $\varphi_{j}(x)$ similarly,
\begin{equation}
\begin{aligned}
  \varphi_{x}
  	=\sum_{n=1}^{\infty} \frac{1}{n!} \int_{y_{1},y_{2},\cdots ,y_{n}} 
  		G_{c}(x,y_{1},y_{2}\cdots,y_{n})
  		\frac{i j_{y_{1}}}{\hbar} \frac{i j_{y_{2}}}{\hbar} \cdots \frac{i j_{y_{n}}}{\hbar}\,.
\end{aligned}\label{classical_scalar_field}
\end{equation}
This relation plays an essential role in deriving the quantum perturbiner expansion, as we will see soon.

The quantum effective action is the functional Legendre transformation of $W[j]$ which interchanges the roles of $j_{x}$ and $\varphi_{x}$,
\begin{equation}
  \Gamma[\varphi]= W[j] - \int_{x} j_{x} \varphi_{x} \,.
\label{}\end{equation}
It is also the generating functional of the 1PI correlation functions. One can show that variation of $\Gamma[\varphi]$ satisfies
\begin{equation}
  \frac{\delta\Gamma[\varphi]}{\delta \varphi_{x}} = \int_{y} \bigg[ \frac{\delta W[j]}{\delta j_{y}} \frac{\delta j_{y}}{\delta\varphi_{x}} - \frac{\delta j_{y}}{\delta\varphi_{x}} \varphi_{y} \bigg] - j_{x} = -j_{x} \,.
\label{}\end{equation}
For the tree-level case, the above equation reduces to the classical EoM. This indicates that the classical field $\varphi_{x}$ can be used to derive the conventional tree-level perturbiner expansion as we will see soon.

For later convenience we further introduce the \emph{descendant fields} $\psi_{x,y}$, $\psi'_{x,y,z}$, $\psi''_{x,y,z,w}, \cdots$ generated by acting multiple functional derivatives on $\varphi_{x}$:
\begin{equation}
\begin{aligned}
  \psi_{x,y} &= \frac{\delta \varphi_{x}}{\delta j_{y}} =  \frac{\delta^{2} W[j]}{\delta j_{x}\delta j_{y}} \,, 
  \qquad
  \psi'_{x,y,z} = \frac{\delta^{2} \varphi_{x}}{\delta j_{y} \delta j_{z}} =\frac{\delta^{3} W[j]}{\delta j_{x} \delta j_{y} \delta j_{z}} \,,
  \\
  \psi''_{x,y,z,w} &= \frac{\delta^{3} \varphi_{x}}{\delta j_{y} \delta j_{z} \delta j_{w}} =\frac{\delta^{4} W[j]}{\delta j_{x} \delta j_{y} \delta j_{z} \delta j_{w}} \,.
\end{aligned}\label{descendant_scalar}
\end{equation}
We define the order of a descendant field as the number of functional derivatives acting on $\varphi(x)$. For instance, $\psi_{x,y}$ and $\psi'_{x,y,z}$ are the first-order and the second-order descendant fields, respectively. We may continue to arbitrarily higher-order descendant fields
\begin{equation}
    \psi^{\scriptsize \overbrace{\prime\prime\cdots \prime}^{n}}_{x,x_{1},x_{2},\cdots,x_{n}} = \frac{\delta^{n} \varphi_{x}}{\delta j_{x_{1}}\delta j_{x_{2}}\cdots \delta j_{x_{n}}} = \frac{\delta^{n+1} W[j]}{\delta j_{x} \delta j_{x_{1}}\delta j_{x_{2}}\cdots \delta j_{x_{n}}}\,.
\label{}\end{equation}
Since the functional derivatives commute with each other, the ordering of the coordinates $x, x_{1}, \cdots$ is irrelevant. We can expand $\varphi_{x}$ and its descendants in $\hbar$,
\begin{equation}
  \varphi_{x} = \sum_{n=0}^{\infty} \bigg(\frac{\hbar}{i}\bigg)^{n} \varphi^{\ord{n}}_{x} \,,
  \qquad
  \psi_{x,y} = \sum_{n=0}^{\infty} \bigg(\frac{\hbar}{i}\bigg)^{n} \psi^{\ord{n}}_{x,y} \,,
  \qquad
  \psi'_{x,y,z} = \sum_{n=0}^{\infty} \bigg(\frac{\hbar}{i}\bigg)^{n} \psi'{}^{\ord{n}}_{x,y,z} \,.
\label{}\end{equation}
%

\subsection{Quantum perturbiner expansion of  $\phi^{4}$-theory}
We now construct the quantum perturbiner expansion generalising the conventional tree-level perturbiner expansion to arbitrary loop-order. To this end, we exploit the quantum effective action formalism reviewed in the previous subsection. Recall that the external source $j_{x}$ is arbitrary, and we are free to choose its form. We identify $j_{x}$ such as to reproduce the LSZ reduction formula out of the expansion of $W[j]$ in \eqref{WJ_scalar},
\begin{equation}
\begin{aligned}
  j_{x} &= \sum_{i=1}^{N}\int_{y_{i}} \,\mathbf{K}_{xy_{i}} e^{-i k_{i}\cdot y_{i}} 
  = \sum_{i=1}^{N} \tilde{\mathbf{K}}(-k_{i}) e^{-i k_{i}\cdot x} \,,
\end{aligned}\label{scalar_external_source}
\end{equation}
where $k_{i}$ are on-shell external momenta for $N$ particles, and $\mathbf{K}_{xy}$ is the inverse of the dressed propagator $\mathbf{D}_{xy}$, satisfying $\int_{y} \mathbf{D}_{x,y} \mathbf{K}_{y,z} = \delta_{x,y}$ or $\tilde{\mathbf{K}}(p) \tilde{\mathbf{D}}(p) = 1$ in momentum space. From \eqref{dressed_propagator_momentum} we can easily read off $\tilde{\mathbf{K}}(p)$ in terms of the self-energy $\Pi(p^{2})$,
\begin{equation}
  \tilde{\mathbf{K}}(p) = p^{2}+m^{2} - \Pi(p^{2}) \,.
\label{}\end{equation}
Thus, we need to compute the 1PI two-point function to define the external source explicitly. Since the self-energy receives loop corrections, we may also expand $j_{x}$ in $\hbar$,
\begin{equation}
  j(x)= j^{\ord{0}}(x)+ \frac{\hbar}{i} j^{\ord{1}}(x) + \bigg(\frac{\hbar}{i}\bigg)^{2} j^{\ord{2}}(x)\,,
\label{}\end{equation}
where
\begin{equation}
\begin{aligned}
  j^{\ord{0}}(x) &= \sum_{i=1}^{N} \big(k_{i}^{2}+m^{2}\big) e^{-ik_{i}\cdot x} \,,
  \\
  j^{\ord{1}}(x) &= - \frac{i}{\hbar}\sum_{i=1}^{N} \Pi^{\ord{1}} e^{-ik_{i}\cdot x} 
  = \frac{\lambda}{2} \sum_{i=1}^{N} \int_{p} \frac{1}{p^{2}+m^{2}} e^{-ik_{i}\cdot x}\,,
\end{aligned}\label{scalar_external_source1}
\end{equation}
and
\begin{equation}
\begin{aligned}
  j^{\ord{2}}_{x} &= - \bigg(\frac{i}{\hbar}\bigg)^{2} \sum_{i=1}^{N} \Pi^{\ord{2}} e^{-ik_{i}\cdot x} 
  \\&
  = - \frac{\lambda^{2}}{4}\sum_{i=1}^{N} \int_{p,q} 
  	\bigg(
  		 \frac{1}{(p^{2}+m^{2})^{2}} \frac{1}{q^{2}+m^{2}} 
  \\&\qquad\qquad\qquad\qquad
  	+ \frac{2}{3} \frac{1}{(p+q-k_{i})^{2}+m^{2}} \frac{1}{p^{2}+m^{2}} \frac{1}{q^{2}+m^{2}}
  	\bigg)e^{-ik_{i}\cdot x}\,.
\end{aligned}\label{current_second}
\end{equation}

If we substitute the $j_{x}$ \eqref{scalar_external_source} into the expansion of $W[j]$ \eqref{WJ_scalar} at $j^{N}$ th order, it reproduces the $N$-point scattering amplitude summed over all permutations of the external particles through the LSZ reduction formula,
\begin{equation}
\begin{aligned}
 \sum_{\text{Perm}[1,\cdots, N]}\mathcal{A}_{k_{1},\cdots,k_{N}} 
 	&= \int_{x_{1},x_{2},\cdots,x_{N}}\frac{\delta^{N} W[j]}{\delta j_{x_{1}} \delta j_{x_{2}}\cdots \delta j_{x_{N}}}\bigg|_{j=0} j_{x_{1}} j_{x_{2}} \cdots j_{x_{N}}\,,
  	\\
  	&=  -i\hbar (2\pi)^{4} \delta^{4}(k_{12\cdots n}) 
  	\sum_{i_{1},i_{2},\cdots,i_{N}} \tilde{G}_{c}(k_{i_{1}},k_{i_{2}},\cdots,k_{i_{N}})\bigg(\frac{i}{\hbar}\bigg)^{N} \prod_{i=1}^{N} \tilde{\mathbf{K}}(-k_{i})
\end{aligned}\label{scalar_LSZ}
\end{equation}
where $k_{12\cdots n} = k_{1} + k_{2} + \cdots + k_{n}$, and $\tilde{G}_{c}(k_{1},k_{2},\cdots,k_{N})$ is the connected $N$-point correlator in momentum space. We now expand the $\varphi_{x}$ \eqref{classical_scalar_field} by substituting $j_{x}$ \eqref{scalar_external_source}. First we define the \emph{quantum off-shell current} $\Phi_{i_{1}\cdots i_{n}}$, which is the amputated correlation function with on-shell momenta $k_{i}$ except for one off-shell leg which is assigned the momentum $-k_{i_{1}\cdots i_{n}} = -\big(k_{i_{1}} + k_{i_{2}} + \cdots + k_{i_{n}}\big)$ by the momentum conservation
\begin{equation}
\begin{aligned}
  \Phi_{i_{1}\cdots i_{n}}
  &=
  	\tilde{G}_{c}(-k_{i_{1}\cdots i_{n}},k_{i_{1}},\cdots,k_{i_{n}})
	\frac{i\tilde{\mathbf{K}}(-k_{i_{1}})}{\hbar} \cdots \frac{i\tilde{\mathbf{K}}(-k_{i_{n}})}{\hbar} \,.
\end{aligned}\label{relation_currents_correlator}
\end{equation}
This leads to our definition of the \emph{quantum perturbiner expansion} of the classical field $\varphi_{x}$ from \eqref{classical_scalar_field}, which generalises the conventional tree-level perturbiner expansion to the quantum level
\begin{equation}
\begin{aligned}
  \varphi_{x} &= \sum_{i=1} \Phi_{i} e^{-ik_{i}\cdot x}+ \sum_{i<j} \Phi_{ij} e^{-ik_{ij}\cdot x} +\cdots+\sum_{i_{1}<i_{2}<\cdots <i_{n}} \Phi_{i_{1}\cdots i_{n}} e^{-ik_{i_{1}\cdots i_{n}}\cdot x}+\cdots\,,
  \\
  &= \sum_{\mathcal{P}} \Phi_{\mathcal{P}} e^{-ik_{\mathcal{P}}\cdot x}\,,
\end{aligned}\label{scalar_perturbiner}
\end{equation}
where $\mathcal{P}, \mathcal{Q}, \mathcal{R} \cdots$ are ordered words which consist of letters, such as $\mathcal{P} = i,j,k,l \cdots $ with $p_{1} < p_{2} < \cdots < p_{|\mathcal{P}|} $, and, which are multi-particle labels. We call the length of the words their `rank' and denoted as $|\mathcal{P}|, |\mathcal{Q}|$, $|\mathcal{R}|$ etc.  The off-shell currents satisfy $\Phi_{\alpha} = \Phi_{\beta}$ for any $\alpha,\beta \in S_{n}$, where $S_{12\cdots n}$ is a permutation group with the set $\{1, 2, \cdots , n \}$. We further require that the off-shell currents with repeated momenta vanish,
\begin{equation}
  \Phi_{i_{1}\cdots j \cdots j\cdots i_{n}} = 0\,.
\label{}\end{equation}
This property excludes currents with higher rank than the number of external particles, $|\mathcal{P}|> N$. The quantum perturbiner generates the quantum off-shell currents, the central object in the off-shell recursion relation.

\begin{figure}[h]
  \includegraphics[scale=0.5]{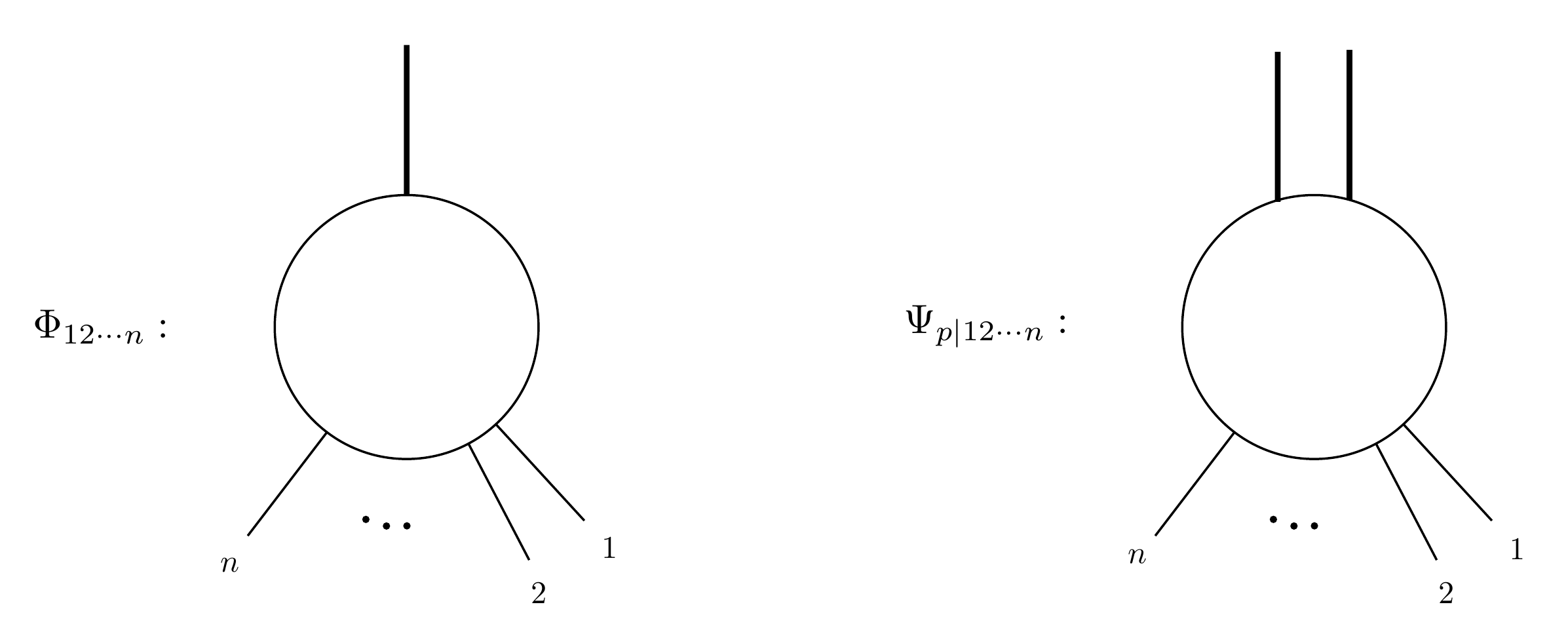}
  \centering
  \caption{Graphical representation of the off-shell currents. The thick lines denote the off-shell legs.}
\label{off_shell_currents}
\end{figure}

We also introduce the quantum perturbiner expansions for the descendant fields $\psi_{x,y}$ and $\psi'_{x,y,z}$ 
\begin{equation}
\begin{aligned}
  \psi_{x,y} &= \int_{p} \Psi_{p|\emptyset} e^{ip\cdot(x-y)} + \sum_{\mathcal{P}} \int_{p} \Psi_{p|\mathcal{P}} e^{ip\cdot(x-y)}e^{-ik_{\mathcal{P}}\cdot x}\,, 
  \\
  \psi'_{x,y,z} &= \sum_{\mathcal{P}} \int_{p,q}\Psi'_{p,q|\mathcal{P}} e^{ip\cdot(x-y)+iq\cdot (x-z)}e^{-ik_{\mathcal{P}}\cdot x}\,, 
\end{aligned}\label{scalar_descendant_perturbiner}
\end{equation}
where $\Psi_{p|\mathcal{P}}$ and $\Psi'_{p,q|\mathcal{P}}$ are the quantum off-shell currents associated with the descendant fields. We will call them as \emph{descendant currents} for $\Phi_{\mathcal{P}}$. Here $p$ and $q$ are off-shell loop momenta, thus $\Psi_{p|\mathcal{P}}$ and $\Psi'_{p,q|\mathcal{P}}$ are intrinsic one-loop and two-loop quantities respectively. Note that the first descendant $\psi_{x,y}$ contains the zero-mode term. The off-shell currents are represented by Feynman diagrams as in Figure \ref{off_shell_currents}. Thus $\Phi_{\mathcal{P}}$ has an off-shell leg, and $n$-th descendant currents have $(n+1)$ off-shell legs.

Since $\tilde{G}_{c}(-k_{i_{1}\cdots i_{n}},k_{i_{1}},\cdots,k_{i_{n}})$ are expanded in powers of $\hbar$, the quantum off-shell currents are also expanded in $\hbar$
\begin{equation}
\begin{aligned}
  \Phi_{\mathcal{P}} = \sum_{n=0}^{\infty} \bigg(\frac{\hbar}{i}\bigg)^{n}\Phi^{(n)}_{\mathcal{P}}\,,
  \qquad
  \Psi_{p|\mathcal{P}} = \sum_{n=0}^{\infty} \bigg(\frac{\hbar}{i}\bigg)^{n}\Psi^{(n)}_{p|\mathcal{P}}\,,
  \qquad
  \Psi'_{p,q|\mathcal{P}} = \sum_{n=0}^{\infty} \bigg(\frac{\hbar}{i}\bigg)^{n}\Psi'^{(n)}_{p,q|\mathcal{P}}\,,\quad \text{etc.}
\end{aligned}\label{hbarExpansion_scalar}
\end{equation}

Analogously to the tree-level case, we can compute loop-level scattering amplitudes from the quantum off-shell currents. The $n$-loop $(N+1)$-point  amplitude $\mathcal{A}^{\ord{n}}(k_{1},\cdots, k_{N+1})$ is given by amputation of the off-shell currents and taking the on-shell limit $(k_{1 \cdots N})^{2} + (m_{\rm phys})^{2} \to 0$,
\begin{equation}
  \mathcal{A}^{\ord{n}}(k_{1},\cdots, k_{N+1}) = \lim_{(k_{1 \cdots N})^{2} \atop \to -(m_{\text{phys}})^{2}} \frac{i}{\hbar}\sum_{p=0}^{n} \tilde{\mathbf{K}}^{\ord{p}}(k_{1\cdots N}) \Phi^{\ord{n-p}}_{1\cdots N}\,.
\label{}\end{equation}
where $\tilde{\mathbf{K}}^{\ord{p}}_{k_{1\cdots N}}$ is the inverse of the dressed propagator in momentum space at $p$-loop order. It is convenient to define the \emph{amputated off-shell current} $\hat{\Phi}^{\ord{n}}_{1\cdots N}$ as
\begin{equation}
  \hat{\Phi}^{\ord{n}}_{1\cdots N} = \frac{i}{\hbar}\sum_{p=0}^{n} \tilde{\mathbf{K}}^{\ord{p}}(k_{1\cdots N}) \Phi^{\ord{n-p}}_{1\cdots N}. 
\label{scalar_amputated_current}\end{equation}
%

\section{The DS equation and The Recursion Relation} \label{Sec:3}
In the conventional tree-level perturbiner method, the off-shell recursion relation is obtained from the classical EoM by substituting the perturbiner expansion into the EoM. Similarly, we construct the quantum off-shell recursion relation by using the quantum perturbiner expansion \eqref{scalar_perturbiner}. To this end, we need to replace the classical EoM with its quantum counterpart. The Dyson--Schwinger (DS) equation is recognized as a quantum analogue of the classical EoM, and it is well-fitted for this purpose. There are several equivalent forms of the DS equation, and here we represent it in terms of $\varphi_{x}$ \cite{Ramond:1981pw,Brown:1992db}.

One may derive the DS equation from the identity for the total functional derivative within a functional integration 
\begin{equation}
\begin{aligned}
  0 &= \int \mathcal{D} \phi \frac{\hbar}{i}\frac{\delta}{\delta \phi_{x}}\ e^{ \frac{i}{\hbar} S[\phi,j] }
  \\
  &= \int \mathcal{D} \phi  \frac{\delta S[\varphi,j]}{\delta \phi_{x}}  e^{ \frac{i}{\hbar} S[\phi,j] } \,.
\end{aligned}\label{}
\end{equation}
If we denote the classical EoM as $\mathcal{F}[\phi] = \frac{\delta S[\phi,0]}{\delta \phi}$, the above relation can be rewritten as
\begin{equation}
  \mathcal{F}\left(\frac{\hbar}{i} \frac{\delta}{\delta j_{x}}\right) Z[j]+ j_{x} Z[j] = 0\,.
\label{}\end{equation}
Using the relation $Z[j] = e^{\frac{i}{\hbar}W[j]}$, we have
\begin{equation}
  e^{-\frac{i}{\hbar} W[j]} \mathcal{F}\left(\frac{\hbar}{i} \frac{\delta}{\delta j_{x}}\right) e^{\frac{i}{\hbar} W[j]} + j_{x} = 0\,.
\label{gen quantum eom}\end{equation}
Then we obtain the DS equation,
\begin{equation}
  \mathcal{F}\left(\varphi_{x}+\frac{\hbar}{i} \frac{\delta}{\delta j_{x}}\right) + j_{x} = 0 \,,
\label{scalar_quantum_eom2}\end{equation}
from the identity
\begin{equation}
  e^{-\frac{i}{\hbar} W[j]} \left(\frac{\hbar}{i} \frac{\delta}{\delta j_{x}}\right) e^{\frac{i}{\hbar} W[j]} = \varphi_{x} + \frac{\hbar}{i} \frac{\delta}{\delta j_{x}}\,.
\label{}\end{equation}

Thus the DS equation is nothing but a deformation of the classical EoM \eqref{phi4_EoM} by $\phi_{x} \to \varphi_{x} + \frac{\hbar}{i}\frac{\delta}{\delta j_{x}}$. If we consider the $\phi^{4}$-theory case, the DS equation is
\begin{equation}
  \int_{y}K(x,y)\varphi_{y} +\frac{\lambda}{3 !}\varphi_{x}^{3} 
  =
  	  j_{x} 
  	- \frac{\lambda}{2} \frac{\hbar}{i} \varphi_{x} \frac{\delta\varphi_{x}}{\delta j_{x}}
  	+ \hbar^{2} \frac{\lambda}{3 !} \frac{\delta^{2}\varphi_{x}}{\delta j_{x}\delta j_{x}}\,.
\label{scalar SD eq}
\end{equation}
We may rewrite the DS equation in terms of the descendant fields \eqref{descendant_scalar},
\begin{equation}
  \int_{y}K(x,y)\varphi_{y} +\frac{\lambda}{3 !}\varphi_{x}^{3} 
  = j_{x} 
  	- \frac{\lambda}{2} \frac{\hbar}{i} \varphi_{x} \psi_{x,x} 
  	+ \hbar^{2} \frac{\lambda}{3 !} \psi'_{x,x,x} \,,
\label{}\end{equation}
or
\begin{equation}
  \varphi_{x} = 
  	\int_{y} D_{xy} 
  		\Big(
  	  		  j_{y} 
  			- \frac{\lambda}{3!} \varphi^{3}_{y}
  		\Big) 
  		+ i\hbar \frac{\lambda}{2} \int_{y}D_{xy} \varphi_{y} \psi_{y,y} 
  		+ \hbar^{2} \frac{\lambda}{3!} \int_{y}D_{xy} \psi'_{y,y,y}\,.
\label{}\end{equation}
Here $\psi_{x,x}$ and $\psi'_{x,x,x}$ are defined by the limits
\begin{equation}
  \psi_{x,x} = \lim_{y\to x} \psi_{x,y}\,, \qquad \psi'_{x,x,x} = \lim_{y\to x\atop z\to x} \psi'_{x,y,z}\,.
\label{1st_SDe}\end{equation}

Our strategy for solving the DS equation is to treat the descendant fields as new independent field variables. Obviously, \eqref{1st_SDe} is not sufficient to obtain a solution because the number of equations is less than the number of undetermined fields. We have to generate field equations for the descendant fields by acting the functional derivative $\frac{\delta}{\delta j_{x}}$ on the DS equation, which we will call  ``descendant equations'',
\begin{equation}
\begin{aligned}	
  \psi_{x,z} &= 
  	  D_{xz}  
  	- \frac{\lambda}{2} \int_{y} D_{xy}\varphi_{y}^{2}\psi_{y,z}
    + i\hbar \frac{\lambda}{2}\int_{y} D_{xy} \big( 
    		  \varphi_{y}\psi'_{y,y,z} 
    		+ \psi_{y,z} \psi_{y,y}
    		\big)
  + \hbar^{2} \frac{\lambda}{3!} \int_{y} D_{xy}  \psi''_{y,y,y,z} \,,
  \\
  \psi'_{x,z,w} &= -\frac{\lambda}{2}\int_{y} D_{xy}
  	\Big(
  		 2\varphi_{y} \psi_{y,w}\psi_{y,z} 
  		+ \varphi_{y}^{2} \psi'_{y,z,w}
  	\Big)  
  \\&\quad
  +i\hbar \frac{\lambda}{2} \int_{y} D_{xy}
  	\Big(
  		  \psi_{y,w}\psi'_{y,y,z} 
	  	+ \phi_{y}\psi''_{y,y,z,w} 
  		+ \psi'_{y,z,w} \psi_{y,y}
  		+ \psi_{y,z} \psi'_{y,y,w}
  	\Big)
  \\&\quad
  +\hbar^{2} \frac{\lambda}{3 !} \int_{y} D_{xy}\psi'''_{y,y,y,z,w}\,.
\end{aligned}\label{scalar_2nd_SDe}
\end{equation}
Again we encounter new undetermined descendant fields such as $\psi''_{x,z}$ and $\psi'''_{x,z,w}$, and we have to generate their field equations to solve them. Apparently, this procedure does not terminate because new descendant fields arise whenever we take functional derivatives on the DS equation. However, we can circumvent the difficulty by expanding the DS equation to a specific loop order. If we substitute the $\hbar$-expansion of all the fields and keep the terms at a fixed order in $\hbar$, we can truncate the new descendant fields because these are higher $\hbar$-order terms in general. 

Once we have the DS equation and its descendants at a specific order in $\hbar$, the recursion relation for the quantum off-shell currents can be derived by substituting the quantum perturbiner expansions \eqref{scalar_perturbiner} and \eqref{scalar_descendant_perturbiner}. We will construct the quantum off-shell recursion relation up to two-loop level and solve the one-loop off-shell currents.

\subsection{Tree level}
Let us start from the tree-level DS equation, which is the same as the classical EoM \eqref{phi4_EoM},
\begin{equation}
  \varphi^{\ord{0}}_{x} = \int_{y} D_{xy} \Big(j^{\ord{0}}_{y} -\frac{\lambda}{3!} \big(\varphi^{\ord{0}}_{y}\big)^{3}\Big)\,.
\label{tree_DS}\end{equation}
If we substitute the perturbiner expansion \eqref{scalar_perturbiner} at $\hbar^{0}$-order,
\begin{equation}
  \varphi^{\ord{0}}_{x} = \sum_{\mathcal{P}} \Phi^{\ord{0}}_{\mathcal{P}} e^{-ik_{\mathcal{P}}\cdot x}\,,
\label{}\end{equation}
we reproduce the conventional tree-level off-shell recursion relation
\begin{equation}
  \Phi^{\ord{0}}_{\mathcal{P}} = -  \frac{\lambda}{3!}\frac{1}{k_{\mathcal{P}}^{2}+m^{2}} \sum_{\mathcal{P}=\mathcal{Q}\cup\mathcal{R}\cup\mathcal{S}} \Phi^{\ord{0}}_{\mathcal{Q}} \Phi^{\ord{0}}_{\mathcal{R}} \Phi^{\ord{0}}_{\mathcal{S}} \,, \qquad |\mathcal{P}|>1 \,,
\label{tree_recursion}\end{equation}
where $\sum_{\mathcal{P}=\mathcal{Q}\cup\mathcal{R}\cup\mathcal{S}}$ means to sum over all possible distributions of the letters of the ordered words $\mathcal{P}$ into non-empty ordered words $\mathcal{Q}$, $\mathcal{R}$ and $\mathcal{S}$. 

To solve the recursion relation, we need to impose the initial condition. In the conventional tree-level perturbiner expansion, it is imposed by hand to reproduce the known scattering amplitudes. On the other hand, we can derive the initial condition from the source term in the DS equation. Since $j^{\ord{0}}(x)$ is proportional to $e^{-i k_{i}\cdot x}$, the initial condition is given by the rank-1 current $\Phi^{\ord{0}}_{i}$. It satisfies 
\begin{equation}
  \sum_{i}\Phi^{\ord{0}}_{i} e^{-ik_{i}\cdot x} = \int_{y} D_{xy} j^{\ord{0}}(y)\,.
\label{}\end{equation}
Substituting $j^{\ord{0}}(x)$ \eqref{scalar_external_source1}, we find the initial condition
\begin{equation}
\begin{aligned}
  \Phi_{i}^{\ord{0}} &= 1 \,,
\end{aligned}\label{}
\end{equation}
which is consistent with the initial condition of the tree-level scalar currents \cite{Mizera:2018jbh}.

\subsection{One loop}
Next we move on to the one-loop level. Substituting the $\hbar$-expansion of $\varphi_{x}$ and $\psi_{x,x}$ \eqref{hbarExpansion_scalar} and keeping $(\hbar)^{1}$-order terms, we have
\begin{equation}
\begin{aligned}
  \varphi^{\ord{1}}_{x} &= \int_{y} D_{xy} \bigg[ j^{\ord{1}}_{y}-\frac{\lambda}{2}\Big( \big(\varphi^{\ord{0}}_{y}\big)^{2}\varphi^{\ord{1}}_{y} +\varphi^{\ord{0}}_{y}\psi^{\ord{0}}_{y,y}\Big)\bigg]\,.
\end{aligned}\label{scalar_one_loop_eom_1}
\end{equation}
It is not sufficient to solve the DS equation due to the presence of $\psi^{\ord{0}}_{y,y}$. Employing the first equation of \eqref{scalar_2nd_SDe} and substituting the $\hbar$-expansion, we derive an additional equation for $\psi^{\ord{0}}_{x,z}$
\begin{equation}
  \psi^{\ord{0}}_{x,z} = D_{xz} -\frac{\lambda}{2} \int_{y} D_{xy}\big(\phi^{\ord{0}}_{y}\big)^{2}\psi^{\ord{0}}_{y,z}\,.
\label{scalar_one_loop_eom_2}\end{equation}
Thus we have a pair of independent equations for the two unknown fields at one-loop. In general, as the loop order increases, the number of equations to be solved increases due to the new descendant fields.

We now construct the one-loop recursion relation. From  \eqref{scalar_perturbiner} and \eqref{scalar_descendant_perturbiner}, the perturbiner expansion for $\varphi_{x}^{\ord{1}}$ and $\psi^{\ord{0}}_{p|\mathcal{P}}$ are given by
\begin{equation}
\begin{aligned}
  \varphi^{\ord{1}}_{x} &= \sum_{\mathcal{P}} \Phi^{\ord{1}}_{\mathcal{P}} e^{-ik_{\mathcal{P}}\cdot x} \,, 
  \\
  \psi^{\ord{0}}_{x,y} &= \int_{p} \Psi^{\ord{0}}_{p|\emptyset} e^{ip\cdot(x-y)} + \sum_{\mathcal{P}} \int_{p} \Psi^{\ord{0}}_{p|\mathcal{P}} e^{ip\cdot(x-y)}e^{-ik_{\mathcal{P}}\cdot x}\,.
\end{aligned}\label{}
\end{equation}
If we substitute the perturbiner expansions into the pair of equations \eqref{scalar_one_loop_eom_1} and \eqref{scalar_one_loop_eom_2}, we obtain the recursion relations at one-loop level
\begin{equation}
\begin{aligned}
  \Phi^{\ord{1}}_{\mathcal{P}} &= 
  	-\frac{\lambda}{2} \frac{1}{(k_{\mathcal{P}})^{2}+m^{2}} \bigg(
  		  \sum_{\mathcal{P}=\mathcal{Q}\cup\mathcal{R}\cup\mathcal{S}}
  			\Phi^{\ord{0}}_{\mathcal{Q}}\Phi^{\ord{0}}_{\mathcal{R}}\Phi^{\ord{1}}_{\mathcal{S}} 
  		+ \sum_{\mathcal{P}=\mathcal{Q}\cup\mathcal{R}} \int_{p}
  			\Phi^{\ord{0}}_{\mathcal{Q}}\Psi^{\ord{0}}_{p|\mathcal{R}}
  		\bigg)\,, \quad \text{for}~ |\mathcal{P}|>1
  \\
  \Psi^{\ord{0}}_{p|\mathcal{P}} &= -\frac{\lambda}{2} \sum_{\mathcal{P}=\mathcal{Q}\cup\mathcal{R}\cup \mathcal{S}}
   		\frac{1}{(p-k_{\mathcal{P}})^{2}+m^{2}} 
   		\Phi^{\ord{0}}_{\mathcal{Q}} \Phi^{\ord{0}}_{\mathcal{R}} \Psi^{\ord{0}}_{p|\mathcal{S}} 
  \qquad 
  \text{for}\ |\mathcal{P}|>0\,.
\end{aligned}\label{scalar_one_loop_recursion_1}
\end{equation}

We have to impose the initial conditions for the quantum off-shell currents to solve the recursions and to reproduce the known scattering amplitudes. Let us consider the initial condition for the descendant current $\Psi^{\ord{0}}_{p|\mathcal{P}}$ first. Since the free propagator $D_{xz}$ on the right-hand side of \eqref{scalar_one_loop_eom_2} does not depend on the external momenta $k_{i}$, it corresponds to the zero-mode, $\Psi^{\ord{0}}_{p|\emptyset}$. Then we can determine $\Psi^{\ord{0}}_{p|\emptyset}$ from the tree-level first descendant equation 
\begin{equation}
  \Psi^{\ord{0}}_{p|\emptyset} = \frac{1}{p^{2}+m^{2}} \,,
\label{psi0_0th}\end{equation}
which is independent of the choice of external source $j_{x}$. We now consider the initial condition for $\Phi^{\ord{1}}_{\mathcal{P}}$. We can determine the lowest-rank current $\Phi^{\ord{1}}_{i}$ from the DS equation at one-loop \eqref{scalar_one_loop_eom_1},
\begin{equation}
\begin{aligned}
  \sum_{i}\Phi^{\ord{1}}_{i} e^{-ik_{i}\cdot x} &=  \int_{y} D_{xy} \bigg[ j^{\ord{1}}_{y}-\frac{\lambda}{2} \sum_{i} \Phi^{\ord{0}}_{i} e^{-ik_{i}\cdot x} \int_{p}\Psi^{\ord{0}}_{p|\emptyset} \bigg]
  \\&
  = \frac{i}{\hbar} \frac{\lambda}{2} \sum_{i}\int_{y} D_{xy} \bigg[\int_{p} \frac{1}{p^{2}+m^{2}} e^{-ik_{i}\cdot y}- \int_{p} \frac{1}{p^{2}+m^{2}} e^{-ik_{i}\cdot y}\bigg]
  \\&
  =0\,.
\end{aligned}\label{}
\end{equation}
Thus the initial condition for the one-loop off-shell current is trivial,
\begin{equation}
  \Phi_{i}^{\ord{1}} = 0 \,.
\label{Phi1_initial}\end{equation}
%

\subsection{Two loops}
Let us consider the DS equation at second order in $\hbar$,
\begin{equation}
  \varphi^{\ord{2}}_{x} =  \int_{y} D_{xy}
  		j^{\ord{2}}_{x}
  		-\frac{\lambda}{2} \int_{y} D_{xy} \bigg(\big(\varphi^{\ord{0}}_{y}\big)^{2} \varphi_{y}^{\ord{2}}
  		+2\varphi^{\ord{0}}_{y} \big(\varphi_{y}^{\ord{1}}\big)^{2}
  		+\big(
  			  \varphi^{\ord{0}}_{y}\psi^{\ord{1}}_{y,y} 
  			+ \varphi^{\ord{1}}_{y}\psi^{\ord{0}}_{y,y}
  		\big) 
  		+ \frac{1}{3} \psi'^{\ord{0}}_{y,y,y}
  	\bigg)\,,
\end{equation}
and the first descendant equation at one-loop order,
\begin{equation}
\begin{aligned}
  \psi_{x,z}^{\ord{1}} &= 
  	- \frac{\lambda}{2} \int_{y}D_{xy}\Big(
  		 2\phi_{y}^{\ord{0}}\phi^{\ord{1}}_{y}\psi^{\ord{0}}_{y,z}
  		+ \big(\phi^{\ord{0}}_{y}\big)^{2} \psi^{\ord{1}}_{y,z}
  		\Big) 
  	- \frac{\lambda}{2} \int_{y}D_{xy}\Big( 
  		  \phi^{\ord{0}}_{y}\psi'^{\ord{0}}_{y,y,z} 
  		+ \psi^{\ord{0}}_{y,y} \psi^{\ord{0}}_{y,z}
  		\Big) \,.
\end{aligned}\label{2nd_SDe}
\end{equation}
In the first equation we encounter the new descendant field $\psi'^{\ord{0}}_{x,y,z}$, and it satisfies the second descendant equation
\begin{equation}
  \psi'^{\ord{0}}_{x,z,w} = -\frac{\lambda}{2} \int_{y}D_{xy} 
  	\Big(
  		2\phi^{\ord{0}}_{y} \psi^{\ord{0}}_{y,z} \psi^{\ord{0}}_{y,w}
  		+\big(\phi_{y}^{\ord{0}}\big)^{2} \psi'^{\ord{0}}_{y,z,w}
  	\Big)\,.
\label{}\end{equation}
The quantum perturbiner expansion for the fields in the two-loop DS equation is
\begin{equation}
\begin{aligned}
  \varphi^{\ord{2}}_{x} &= \sum_{\mathcal{P}} \Phi^{\ord{2}}_{\mathcal{P}} e^{-ik_{\mathcal{P}}\cdot x}\,,
  \\
  \psi^{\ord{1}}_{x,y} &= \int_{p} \Psi^{\ord{1}}_{p|\emptyset} e^{ip\cdot(x-y)} + \sum_{\mathcal{P}} \int_{p} \Psi^{\ord{1}}_{p|\mathcal{P}} e^{ip\cdot(x-y)}e^{-ik_{\mathcal{P}}\cdot x}\,, 
  \\
  \psi'^{\ord{0}}_{x,y,z} &= \sum_{\mathcal{P}} \int_{p,q}\Psi^{\ord{0}}_{p,q|\mathcal{P}} e^{ip\cdot(x-y)+iq\cdot (x-z)}e^{-ik_{\mathcal{P}}\cdot x}\,.
\end{aligned}\label{two-loop_perturbiner}
\end{equation}

We now construct the recursion relation by substituting \eqref{two-loop_perturbiner} into the two-loop DS equation and its descendants,
 \begin{equation}
\begin{aligned}
  \Phi^{\ord{2}}_{\mathcal{P}} 
  &= -\frac{\lambda}{2} \frac{1}{k_{\mathcal{P}}^{2}+m^{2}} 
  	\bigg(
  		\sum_{\mathcal{P}=\mathcal{Q}\cup\mathcal{R}\cup\mathcal{S}} 
  			\Big(
  			  \Phi^{\ord{0}}_{\mathcal{Q}} \Phi^{\ord{0}}_{\mathcal{R}} \Phi^{\ord{2}}_{\mathcal{S}} 
  				+2\Phi^{\ord{0}}_{\mathcal{Q}} \Phi^{\ord{1}}_{\mathcal{R}} \Phi^{\ord{1}}_{\mathcal{S}}
  			\Big)
  \\&\qquad\qquad\qquad\quad
  		+ \sum_{\mathcal{P}=\mathcal{Q}\cup\mathcal{R}} \int_{p} 
  			\Big(
  			  \Phi^{\ord{0}}_{\mathcal{Q}} \Psi^{\ord{1}}_{p|\mathcal{R}}
  			+ \Phi^{\ord{1}}_{\mathcal{Q}} \Psi^{\ord{0}}_{p|\mathcal{R}}
  			\Big)
  		+ \frac{1}{3} \int_{p,q}\Psi'^{\ord{0}}_{p,q|\mathcal{P}}
  	\bigg)\,, 
  \\  
	\Psi^{\ord{1}}_{p|\mathcal{P}} &= 
  	-\frac{\lambda}{2} \frac{1}{(p-k_{\mathcal{P}})^{2}+m^{2}} 
  	\sum_{\mathcal{P}=\mathcal{Q}\cup\mathcal{R}\cup\mathcal{S}}
  		\Big( 
  			 2\Phi^{\ord{0}}_{\mathcal{Q}} \Phi^{\ord{1}}_{\mathcal{R}} \Psi^{\ord{0}}_{p|\mathcal{S}} 
  			+ \Phi^{\ord{0}}_{\mathcal{Q}} \Phi^{{\ord{0}}}_{\mathcal{R}} \Psi^{\ord{1}}_{p|\mathcal{S}}
  		\Big)
  	\\&\quad
  	- \frac{\lambda}{2} \frac{1}{(p-k_{\mathcal{P}})^{2}+m^{2}} 
 	  \sum_{\mathcal{P}=\mathcal{Q}\cup\mathcal{R}}\int_{q} 
  		\Big(
  		  \Phi^{\ord{0}}_{\mathcal{Q}}\Psi'^{\ord{0}}_{p,q|\mathcal{R}}
  		+ \Psi^{\ord{0}}_{p|\mathcal{Q}} \Psi^{\ord{0}}_{q|\mathcal{R}}
  		\Big)\,,
  \\  
  \Psi'^{\ord{0}}_{p,q|\mathcal{P}} &= -\frac{\lambda}{2} \frac{1}{(p+q-k_{\mathcal{P}})^{2}+m^{2}}
  	\sum_{\mathcal{P}=\mathcal{Q}\cup\mathcal{R}\cup\mathcal{S}}\Big(
  		 2\Phi^{\ord{0}}_{\mathcal{Q}} \Psi_{p|\mathcal{R}}^{\ord{0}} \Psi^{\ord{0}}_{q|\mathcal{S}} 
  		+ \Phi_{\mathcal{Q}}^{\ord{0}} \Phi_{\mathcal{R}}^{\ord{0}} \Psi'^{\ord{0}}_{p,q|\mathcal{S}}
  	\Big) \,.
\end{aligned}\label{scalar_two_loop_recursion2}
\end{equation}

Let us consider the initial conditions for each off-shell current. As discussed for the one-loop level, we can derive the initial conditions from the external source in the DS equation. The rank-1 current $\Phi^{\ord{2}}_{i}$ satisfies
\begin{equation}
\begin{aligned}
  \sum_{i} \Phi^{\ord{2}}_{i} e^{-ik_i\cdot x} &= \int_{y} D_{xy} j^{\ord{2}}_{y}-\frac{\lambda}{2}\int_{y} \sum_{i} D_{xy} \bigg(\Phi^{\ord{0}}_{i}\int_{p} \Psi^{\ord{1}}_{p|\emptyset}
  	+ \frac{1}{3}\int_{p,q}\Psi'^{\ord{0}}_{p,q|i}\bigg) e^{-ik_{i}\cdot y} \,.
\end{aligned}\label{}
\end{equation}
The descendant current $\Psi^{\ord{1}}_{p|\emptyset}$ satisfies the zero-mode sector of the first descendant of the DS equation,
\begin{equation}
\begin{aligned}
  \Psi^{\ord{1}}_{p|\emptyset} &= - \frac{\lambda}{2} \frac{1}{p^{2}+m^{2}} 
  	\int_{q} \Psi^{\ord{0}}_{p|\emptyset} \Psi^{\ord{0}}_{q|\emptyset} \,,
  	\\
  	&= - \frac{\lambda}{2} \bigg(\frac{1}{p^{2}+m^{2}}\bigg)^{2}
  	\int_{q} \frac{1}{q^{2}+m^{2}} \,.
\end{aligned}\label{}
\end{equation}
and 
\begin{equation}
\begin{aligned}
  \Psi'^{\ord{0}}_{p,q|i} &= -\frac{\lambda}{2} \frac{2}{(p+q-k_{i})^{2}+m^{2}}
  		 \Phi^{\ord{0}}_{i} \Psi_{p|\emptyset}^{\ord{0}} \Psi^{\ord{0}}_{q|\emptyset}
  \\
  &= -\frac{\lambda}{2} \frac{2}{(p+q-k_{i})^{2}+m^{2}}
  		\frac{1}{p^{2}+m^{2}} \frac{1}{q^{2}+m^{2}} \,.
\end{aligned}\label{}
\end{equation}
From the explicit form of $j^{\ord{2}}_{x}$ in \eqref{current_second}, we can show that the initial condition for the two-loop off-shell current is trivial,
\begin{equation}
  \Phi^{\ord{2}}_{i} = 0\,.
\label{}\end{equation}
We may expect that this is a general property of the initial conditions of the loop-level currents,
\begin{equation}
  \Phi^{\ord{n}}_{i} = 0\,, \qquad \text{for}~ n>1 \,.
\label{}\end{equation}

In the next subsection, we will focus on solving the one-loop recursion relation and determining the off-shell currents $\Phi^{\ord{1}}_{\mathcal{P}}$, $\Psi^{\ord{1}}_{\mathcal{P}}$ and $\Psi'^{\ord{1}}_{\mathcal{P}}$.

\subsection{Solving the Recursion Relation: One loop}
\begin{figure}[h]
  \includegraphics[scale=0.5]{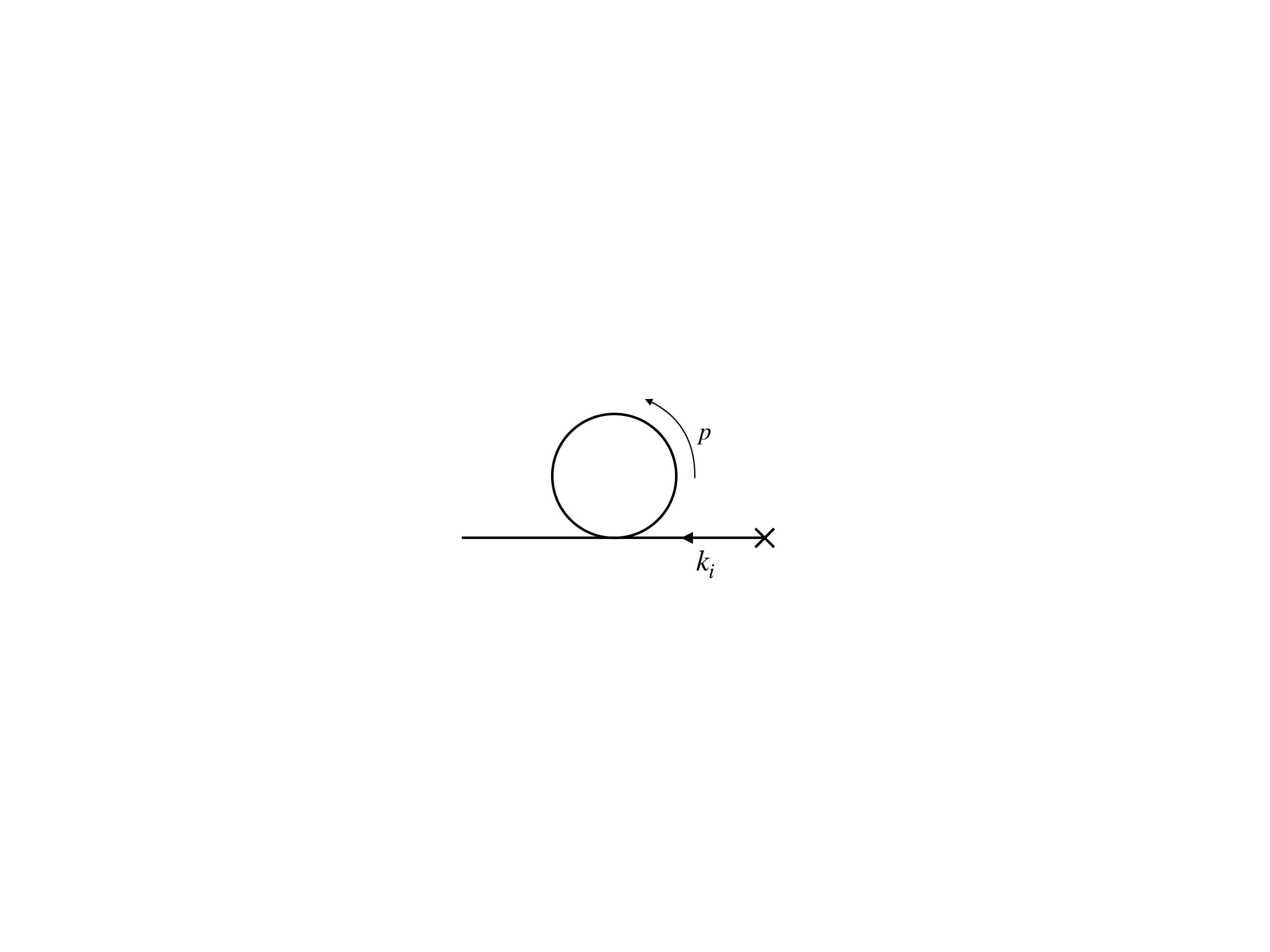}
  \centering
  \caption{One-loop rank-3 amputated off-shell currents. Here the thick line is the amputated off-shell leg, and the ellipsis represents the other independent permutations of $i,j$ and $k$. }
\label{sclar_1loop_rank3}
\end{figure}

We now compute the one-loop off-shell currents $\Phi^{\ord{1}}_{\mathcal{P}}$ by solving the one-loop recursion relations iteratively. Since the one-loop recursion relations involve loop integrals, these require regularisation and renormalisation to obtain well-defined finite results. Here we will not evaluate the loop integrations, but we will show the consistency with the known results from Feynman diagram.
\paragraph{$\bullet$ Rank-1}
~\\
As we have shown that the one-loop initial condition is trivial \eqref{Phi1_initial}, the one-loop rank-1 current vanishes, $\Phi_{i}^{\ord{1}} = 0$.
\paragraph{$\bullet$ Rank-2}
~\\
From the one-loop recursion relation \eqref{scalar_one_loop_recursion_1}, the one-loop rank-2 current $\Phi^{\ord{1}}_{ij}$ satisfies
\begin{equation}
  \Phi^{\ord{1}}_{ij} = -\frac{\lambda}{2} \frac{1}{k^{2}_{ij}+m^{2}} \sum_{i,j} \Big(\Phi^{\ord{0}}_{ij} \int_{p}\Psi^{\ord{0}}_{p|\emptyset} + 2 \Phi^{\ord{0}}_{i} \int_{p}\Psi^{\ord{0}}_{p|j}\Big)\,.
\label{}\end{equation}
Note that $\Phi^{\ord{0}}_{ij} = 0$ because the nontrivial tree-level recursion \eqref{tree_recursion} starts from the rank-3, and the first term on the righthand side does not contribute. It is also straightforward to show that $\Psi^{\ord{0}}_{p|i}= 0$ using the recursion relations \eqref{scalar_one_loop_recursion_1} and \eqref{scalar_one_loop_recursion_2}, thus we have
\begin{equation}
  \Phi^{\ord{1}}_{ij}= 0\,.
\label{}\end{equation}
This result is consistent with the fact that there is no one-loop three-point function in $\phi^{4}$-theory.
\paragraph{$\bullet$ Rank-3}
~\\
We can obtain the one-loop rank-3 current from the recursion relation \eqref{scalar_one_loop_recursion_1},
\begin{equation}
  \Phi^{\ord{1}}_{ijk} = -\frac{\lambda}{2} \frac{1}{(k_{ijk})^{2}+m^{2}} \sum_{\text{Perm}[i,j,k]}
  	\bigg(
  		  \frac{1}{2} \int_{p}\Phi^{\ord{0}}_{i}\Psi^{\ord{0}}_{p|jk} 
  		+ \frac{1}{3!} \int_{p}\Phi^{\ord{0}}_{ijk}\Psi^{\ord{0}}_{p|\emptyset}\bigg)\,.
\label{}\end{equation}
First we need to determine $\Phi^{\ord{0}}_{ijk}$ and $\Psi^{\ord{0}}_{p|ij}$. From the tree-level recursion relation \eqref{tree_recursion},  we can solve $\Phi^{\ord{0}}_{ijk}$ as
\begin{equation}
\begin{aligned}
  \Phi^{\ord{0}}_{ijk} 
  &= - \frac{\lambda}{3!} \frac{1}{k_{ijk}^{2}+m^{2}} \sum_{\text{Perm}[i,j,k]} \Phi^{(0)}_{i} \Phi^{(0)}_{j} \Phi^{(0)}_{k} 
  \\
  &= -\frac{\lambda}{k_{ijk}^{2}+m^{2}} \,.
\end{aligned}\label{}
\end{equation}
Next, we consider $\Psi^{\ord{0}}_{p|ij}$ using \eqref{scalar_one_loop_recursion_1},
\begin{equation}
\begin{aligned}
  \Psi^{\ord{0}}_{p|ij} 
  &= -\lambda \frac{1}{(p-k_{ij})^{2}+m^{2}}\Phi^{\ord{0}}_{i} \Phi^{\ord{0}}_{j} \Psi^{\ord{0}}_{p|\emptyset} 
  \\
  &= -\lambda \frac{1}{(p-k_{ij})^{2}+m^{2}} \frac{1}{p^{2}+m^{2}}\,.
\end{aligned}\label{}
\end{equation}
If we collect all the ingredients, the one-loop rank-3 current reduces to
\begin{equation}
\begin{aligned}
  \Phi^{\ord{1}}_{ijk} = \frac{\lambda^{2}}{2} \frac{1}{k_{ijk}^{2}+m^{2}}\bigg(&  \frac{1}{k_{ijk}^{2}+m^{2}}\int_{p} \frac{1}{p^{2}+m^{2}}
  \\
  &+ \int_{p}  \frac{1}{\big(p^{2}+m^{2}\big)\big((p-k_{ij})^{2}+m^{2}\big)} + \big(ij \to jk \to ik\big)
  \bigg)\,.
\end{aligned}\label{}
\end{equation}
The corresponding amputated off-shell current $\hat{\Phi}^{\ord{1}}_{ijk}$ is given by
\begin{equation}
  \hat{\Phi}^{\ord{1}}_{ijk} = \frac{\lambda^{2}}{2} \int_{p} \frac{1}{\big(p^{2}+m^{2}\big)\big((p-k_{ij})^{2}+m^{2}\big)} + \big(ij \to jk \to ik\big)\,.
\label{}\end{equation}
The diagrammatical representation of the rank-3 amputated current is given in Figure \ref{sclar_1loop_rank3}. It reproduces the one-loop 4-point scattering amplitude for $\phi^{4}-$theory.

\begin{figure}[t]
  \includegraphics[scale=0.5]{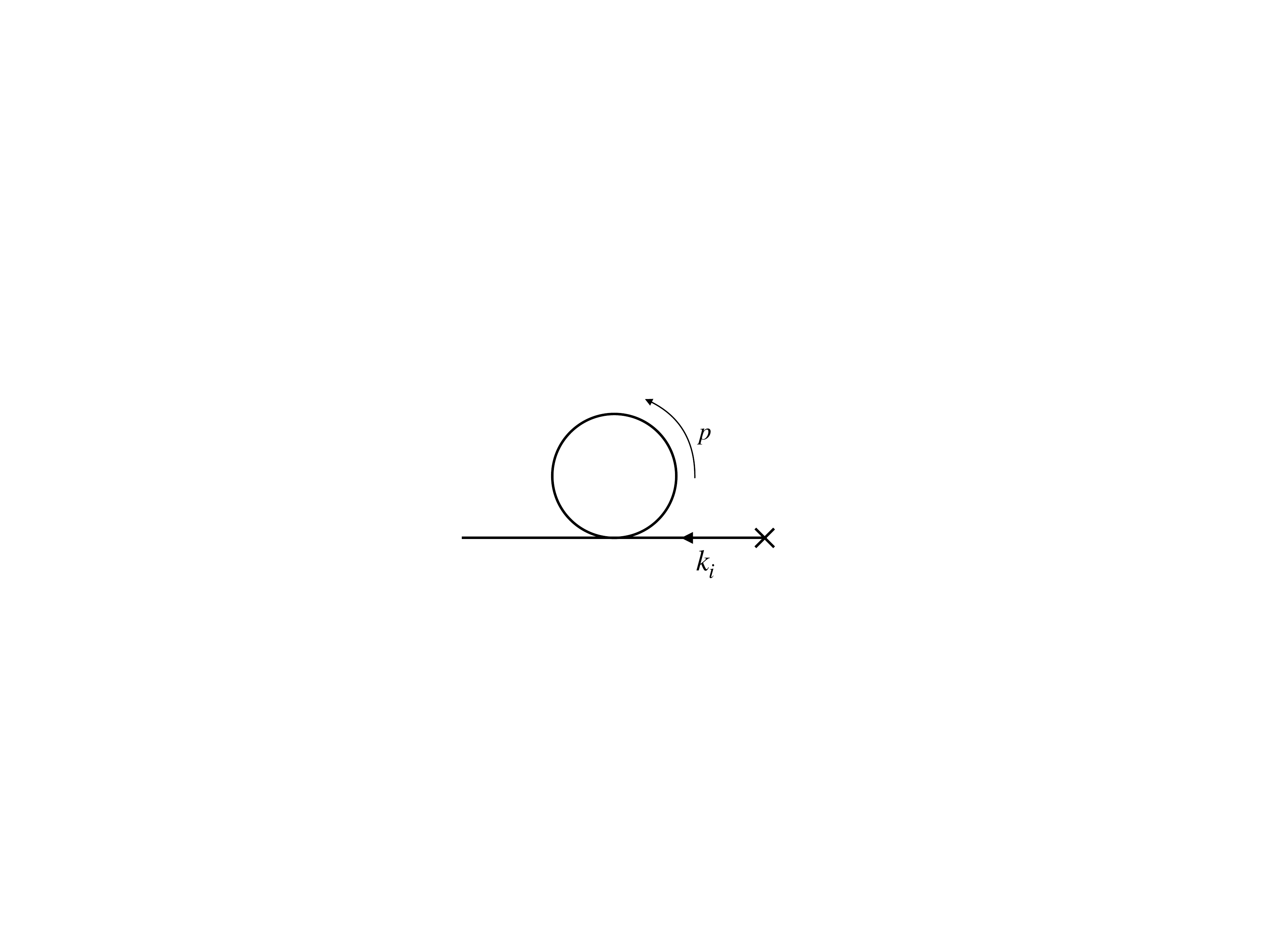}
  \centering
  \caption{One-loop six-point amplitude from the one-loop rank-5 off-shell currents.}
\label{sclar_1loop_rank5}
\end{figure}

\paragraph{$\bullet$ Rank-4}
~\\
From the recursion relation \eqref{scalar_one_loop_recursion_1}, $\Phi^{\ord{1}}_{ijkl}$ is given by
\begin{equation}
\begin{aligned}
  \Phi^{\ord{1}}_{ijkl} = -\frac{\lambda}{2} \frac{1}{k_{ijkl}^{2}+m^{2}} &\sum_{\text{Perm}[ijkl]} 
  	\bigg[
  		  \frac{1}{2} \Phi^{\ord{0}}_{i} \Phi^{\ord{0}}_{j} \Phi^{\ord{1}}_{kl} 
  		+ \Phi^{\ord{0}}_{ij} \Phi^{\ord{0}}_{k} \Phi^{\ord{1}}_{l} 
  		\\&~~
  		+\int_{p} \Big(
  			  \frac{1}{4!}\Phi^{\ord{0}}_{ijkl} \Psi^{\ord{0}}_{p|\emptyset} 
  			+ \frac{1}{3!}\Phi^{\ord{0}}_{ijk} \Psi^{\ord{0}}_{p|l}
  			+ \frac{2}{(2!)^2}\Phi^{\ord{0}}_{ij} \Psi^{\ord{0}}_{p|kl} 
  			+ \frac{1}{3!}\Phi^{\ord{0}}_{i} \Psi^{\ord{0}}_{p|jkl} 
  			\Big)
  	\bigg]
\end{aligned}\label{}
\end{equation}
Using the previous results, $\Phi_{ij}^{\ord{0}} = \Phi_{ij}^{\ord{1}} = \Psi^{\ord{0}}_{p|i} =0$, and the remaining unknown current is $\Psi^{\ord{0}}_{p|ijk}$. From the recursion relation, it is given by
\begin{equation}
  \Psi^{\ord{0}}_{p|ijk} 
  = \frac{\lambda}{2} \frac{1}{(p-k_{ijk})^{2}+m^{2}} \sum_{\text{perm}[ijk]} \Phi^{\ord{0}}_{i} \Phi^{\ord{0}}_{j} \Psi^{\ord{0}}_{p|k} 
  = 0\,.
\label{}\end{equation}
Since $\Psi^{\ord{0}}_{p|i} =0$, $\Psi^{\ord{0}}_{p|ijk}$ also vanishes. Thus, this shows that the one-loop rank-4 current $\Phi^{\ord{1}}_{ijkl}$ and the one-loop 5-point amplitude vanish,
\begin{equation}
  \Phi^{\ord{1}}_{ijkl} = 0\,.
\label{}\end{equation}
\paragraph{$\bullet$ Rank-5}
~\\
The one-loop rank-5 current $\Phi^{\ord{1}}_{i_{1}i_{2}\cdots i_{5}}$ satisfies the recursion relation
\begin{equation}
\begin{aligned}
  \Phi^{\ord{1}}_{i_{1}\cdots i_{5}} = -\frac{\lambda}{2} \frac{1}{k^{2}_{i_{1}\cdots i_{5}}+m^{2}} &\sum_{\text{Perm}[i_{1},\cdots,i_{5}]} 
  	\Bigg[ \
  		  \frac{1}{3!} \Phi^{\ord{0}}_{i_{1}}\Phi^{\ord{0}}_{i_{2}}\Phi^{\ord{1}}_{i_{3}i_{4}i_{5}} 
  \\&\quad
  		+\int_{p} \bigg(\frac{1}{5!} \Phi^{\ord{0}}_{i_{1}\cdots i_{5}}\Psi^{\ord{0}}_{p|\emptyset} 
  		+ \frac{1}{2!}\frac{1}{3!} \Phi^{\ord{0}}_{i_{1}i_{2}i_{3}}\Psi^{\ord{0}}_{p|i_{4}i_{5}}
  		+ \frac{1}{4!} \Phi^{\ord{0}}_{i_{1}}\Psi^{\ord{0}}_{p|i_{2}i_{3}i_{4}i_{5}} \bigg)
  	\Bigg]
\end{aligned}\label{}
\end{equation}
We omitted terms containing $\Phi^{\ord{0}}_{ij}$, $\Phi^{\ord{0}}_{ijkl}$ and $\Psi^{\ord{0}}_{p|i}$ because these are trivial, as we have shown before. 
The rank-4 descendant current $\Psi^{\ord{0}}_{p|ijkl}$ can be obtained from the recursion relation
\begin{equation}
\begin{aligned}
  \Psi^{\ord{0}}_{p|ijkl} &= \frac{\lambda}{2} \frac{1}{(p-k_{ijkl})^{2}+m^{2}} \sum_{\text{Perm}[ijkl]}
  	\bigg(
  		  \frac{1}{2!}\Phi^{(0)}_{i}\Phi^{(0)}_{j}\Psi^{(0)}_{p|kl}
  		+ \frac{1}{3!}\Phi^{(0)}_{i}\Phi^{(0)}_{jkl} \Psi^{(0)}_{p|\emptyset}
  	\bigg)
  \\
  &= - \frac{\lambda^{2}}{4} \frac{1}{(p-k_{ijkl})^{2}+m^{2}} \sum_{\text{Perm}[ijkl]} \frac{1}{p^{2}+m^{2}}
  	\bigg[ \frac{1}{(p-k_{kl})^{2}+m^{2}} +\frac{1}{3} \frac{1}{k_{jkl}^{2}+m^{2}}\bigg] \,.
\end{aligned}\label{}
\end{equation}
The amputated off-shell current at this order $\hat{\Phi}^{\ord{1}}_{i_{1}\cdots i_{5}}$ is given by
\begin{equation}
\begin{aligned}
  \hat{\Phi}^{\ord{1}}_{i_{1}\cdots i_{5}} = -\frac{\lambda}{2} & \sum_{\text{Perm}[i_{1},\cdots,i_{5}]} 
  	\Bigg[ \
  		  \frac{1}{3!} \Phi^{\ord{1}}_{i_{3}i_{4}i_{5}} 
  		+\int_{p} \bigg(
  		 \frac{1}{2\cdot 3!} \Phi^{\ord{0}}_{i_{1}i_{2}i_{3}}\Psi^{\ord{0}}_{p|i_{4}i_{5}}
  		+ \frac{1}{4!} \Psi^{\ord{0}}_{p|i_{2}i_{3}i_{4}i_{5}} \bigg)
  	\Bigg]\,.
\end{aligned}\label{}
\end{equation}
Collecting all the ingredients, we can represent the rank-5 current diagrammatically in Figure \ref{sclar_1loop_rank5}.
It precisely reproduces the one-loop 6-point scattering amplitude.

\section{Quantum Perturbiner Method for Correlation Functions}\label{Sec:4}
So far, we have constructed the quantum perturbiner method for computing loop-level scattering amplitudes. When we construct the perturbiner expansion, we have chosen an external source that reproduces the LSZ reduction formula from the connected generating functional $W[j]$. However, the choice of the external source is not unique at all, and we may replace it depending on our purpose. The associated off-shell current describes different quantities as we modify the external source.

In this section, we construct the quantum perturbiner method for computing connected correlation functions, $G_{c}(x_{1},\cdots x_{n})$, instead of scattering amplitudes. To this end, we replace the external source $j(x)$ defined in \eqref{scalar_external_source} with a new current $\check{j}(x)$,
\begin{equation}
\begin{aligned}
  \check{j}(x) &= \frac{\hbar}{i}\sum_{i=1}^{N}\int_{y_{i}}\, \delta^{4}(x-y_{i}) e^{-i k_{i}\cdot y_{i}}
  \\
  &= \frac{\hbar}{i} \sum_{i=1}^{N} e^{-ik_{i}\cdot x}\,,
\end{aligned}\label{source_correlator}
\end{equation}
where $k_{i}$ are the external momenta without an on-shell condition. We have dropped the inverse propagator $\mathbf{K}_{xy}$ in the external source because we do not have to amputate the external legs when computing correlation functions.

Most of the results in Section \ref{Sec:2} still hold. The classical field $\check{\varphi}_{x}$ with respect to the new external source is given by
\begin{equation}
  \check{\varphi}_{x} = \frac{\delta W[\check{j}]}{\delta \check{j}(x)}\,,
\label{}\end{equation}
and the classical field is expanded as
\begin{equation}
\begin{aligned}
  \check{\varphi}_{x} &= \sum_{i=1} \check{\Phi}_{i} e^{-ik_{i}\cdot x}+ \sum_{i<j} \check{\Phi}_{ij} e^{-ik_{ij}\cdot x} +\cdots+\sum_{i_{1}<i_{2}<\cdots <i_{N}} \check{\Phi}_{i_{1}\cdots i_{N}} e^{-ik_{i_{1}\cdots i_{N}}\cdot x}
  \\&
  = \sum_{\mathcal{P}} \check{\Phi}_{\mathcal{P}} e^{-ik_{\mathcal{P}}\cdot x}\,.
\end{aligned}\label{}
\end{equation}
Then the rank-$n$ off-shell currents in the perturbiner expansion of $\varphi_{x}$ provide the momentum space $(n+1)$-point correlation functions
\begin{equation}
\begin{aligned}
  \check{\Phi}_{i_{1}i_{2}\cdots i_{n}} &= \tilde{G}_{c}(-k_{i_{1}i_{2}\cdots i_{n}},k_{i_{1}},k_{i_{2}},\cdots,k_{i_{n}}) \,.
\end{aligned}\label{}
\end{equation}
Thus we can identify the off-shell currents with correlation functions.

\subsection{Recursion relation and its solution}
Since the form of the perturbiner expansion and the DS equation are invariant with respect to the choice of external source, the off-shell recursion relations are the same as the previous results in Section \ref{Sec:3}. The only difference is the initial condition because it directly depends on the choice of external source explicitly. Here we will derive the modified initial condition, $\check{\Phi}^{\ord{n}}_{i}$, by solving the DS equation with $\check{j}_{x}$. Further, we will solve the recursion relation and determine the correlation functions up to the two-loop level.

The tree-level recursion relation, which is the same as \eqref{tree_recursion}, is given by
\begin{equation}
  \check{\Phi}^{\ord{0}}_{\mathcal{P}} = -  \frac{\lambda}{3!}\frac{1}{k_{\mathcal{P}}^{2}+m^{2}} \sum_{\mathcal{P}=\mathcal{Q}\cup\mathcal{R}\cup\mathcal{S}} \check{\Phi}^{\ord{0}}_{\mathcal{Q}} \check{\Phi}^{\ord{0}}_{\mathcal{R}} \check{\Phi}^{\ord{0}}_{\mathcal{S}} \qquad \text{for} ~|\mathcal{P}|>1 \,.
\label{}\end{equation}
The initial condition depends on the choice of the external source. From the DS equation, the rank-1 current satisfies
\begin{equation}
  \sum_{i}\check{\Phi}^{\ord{0}}_{i} e^{-k_{i}\cdot x} = \int_{y} D_{xy} \check{j}^{\ord{0}}(y) 
  = \frac{\hbar}{i} \sum_{i} \frac{e^{-ik_{i}\cdot x}}{k_{i}^{2}+m^{2}}\,,
\label{}\end{equation}
where the $k_{i}$ are off-shell momenta. It is straightforward to read off the initial condition 
\begin{equation}
  \check{\Phi}_{i} = \frac{\hbar}{i} \frac{1}{k_{i}^{2}+m^{2}}\,.
\label{}\end{equation}
This is the tree-level two-point function. 

We can solve the recursion relation and derive the tree-level correlation functions in momentum space iteratively. We list the tree-level currents up to rank-5
\begin{equation}
\begin{aligned}
  \check{\Phi}_{i} &= \frac{\hbar}{i} \frac{1}{k_{i}^{2}+m^{2}} \,,
  \\
  \check{\Phi}_{ij} &= 0\,,
  \\
  \check{\Phi}_{i_{1}i_{2}i_{3}} &= - i \hbar^{3} \lambda \frac{1}{k_{i_{1}i_{2}i_{3}}^{2}+m^{2}} \prod_{m=1}^{3} \frac{1}{k_{i_{m}}^{2}+m^{2}}\,,
  \\
  \check{\Phi}_{i_{1}i_{2}i_{3}i_{4}} &= 0\,,
  \\
  \check{\Phi}_{i_{1}i_{2}i_{3}i_{4}i_{5}} &= - \hbar^{5}\frac{\lambda^{2}}{(3!)^{2}} \frac{1}{(k_{i_{1} \cdots i_{5}})^{2}+ m^{2}} \sum_{\text{Perm}[i_{1},\cdots,i_{5}]} \frac{1}{k_{i_{1}i_{2}i_{3}}^{2}+m^{2}} \prod_{m=1}^{5} \frac{1}{k_{i_{m}}^{2}+m^{2}} \,,
\end{aligned}\label{}
\end{equation}
and the result is the same as the tree-level connected correlation functions.

\subsection{One-loop level}
\begin{figure}[t]
  \includegraphics[scale=0.53]{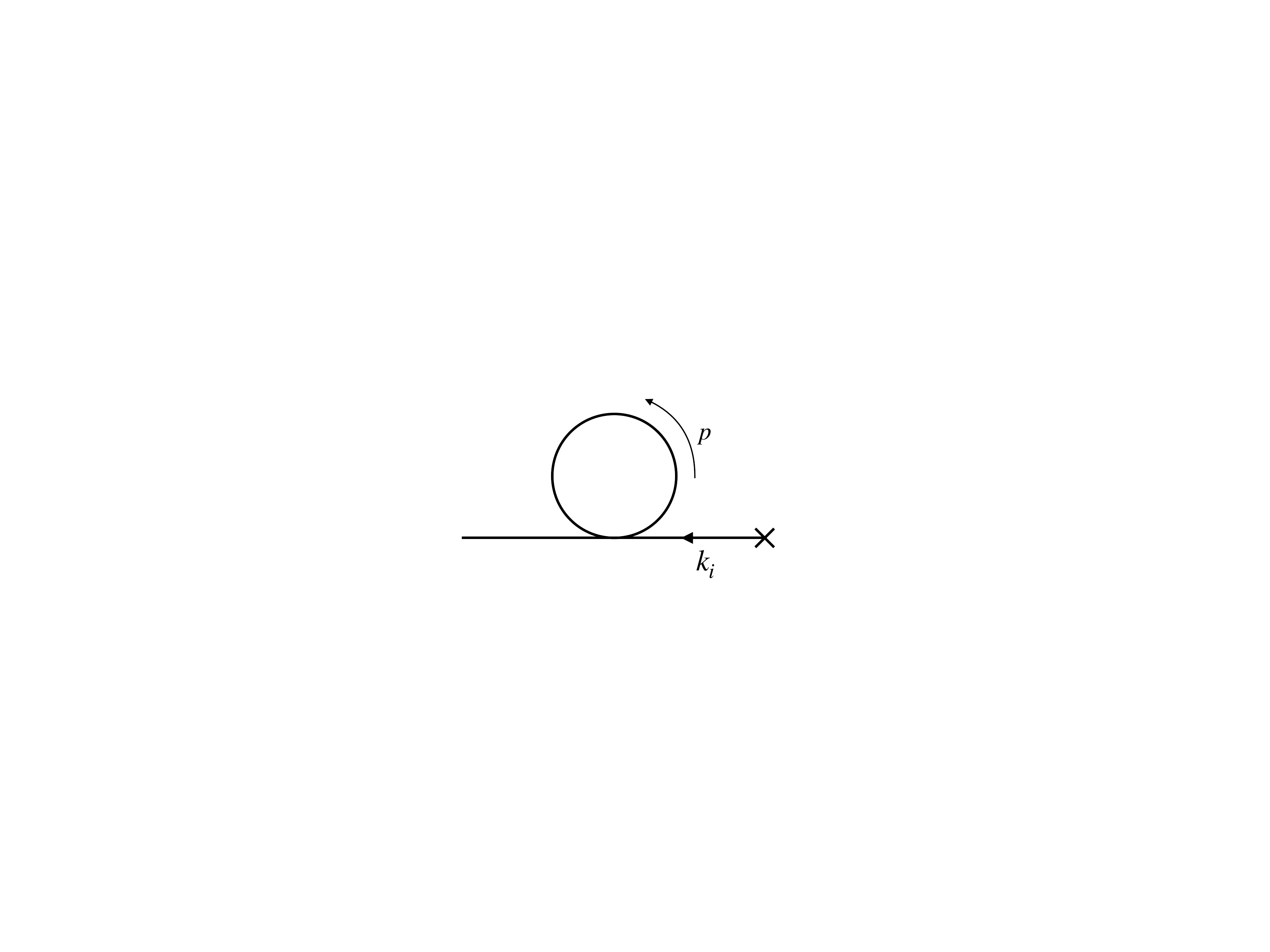}
  \centering
  \caption{1-loop rank-3 off-shell currents that correspond to the 4-point correlation function.}
\label{sclar_1loop_rank3_correlation}
\end{figure}

The recursion relations for one-loop currents $\check{\Phi}^{\ord{1}}_{\mathcal{P}}$ and $\check{\Psi}^{\ord{0}}_{p|\mathcal{P}}$ are the same as the previous result \eqref{scalar_one_loop_recursion_1} ,
\begin{equation}
\begin{aligned}
  \check{\Phi}^{\ord{1}}_{\mathcal{P}} &= 
  	-\frac{\lambda}{2} \frac{1}{(k_{\mathcal{P}})^{2}+m^{2}} \bigg(
  		  \sum_{\mathcal{P}=\mathcal{Q}\cup\mathcal{R}\cup\mathcal{S}}
  			\check{\Phi}^{\ord{0}}_{\mathcal{Q}} \check{\Phi}^{\ord{0}}_{\mathcal{R}} \check{\Phi}^{\ord{1}}_{\mathcal{S}} 
  		+ \sum_{\mathcal{P}=\mathcal{Q}\cup\mathcal{R}} \int_{p}
  			\check{\Phi}^{\ord{0}}_{\mathcal{Q}} \check{\Psi}^{\ord{0}}_{p|\mathcal{R}}
  		\bigg) \quad \text{for}~ |\mathcal{P}|>1\,,
  \\
  \check{\Psi}^{\ord{0}}_{p|\mathcal{P}} &= \frac{\lambda}{2} \sum_{\mathcal{P}=\mathcal{Q}\cup\mathcal{R}\cup \mathcal{S}}
   		\frac{1}{(p-k_{\mathcal{P}})^{2}+m^{2}} 
   		\check{\Phi}^{\ord{0}}_{\mathcal{Q}} \check{\Phi}^{\ord{0}}_{\mathcal{R}} \check{\Psi}^{\ord{0}}_{p|\mathcal{S}} 
  \qquad 
  \text{for}\ |\mathcal{P}|>0\,.
\end{aligned}\label{scalar_one_loop_recursion_2}
\end{equation}
First, the initial condition for the descendant current $\check{\Psi}^{\ord{0}}_{p|\emptyset}$ is identical with the previous one \eqref{psi0_0th} because it is independent of the form of the external source,
\begin{equation}
  \check{\Psi}^{(0)}_{p|\emptyset} = \frac{1}{p^{2}+m^{2}} \,.
\label{}\end{equation}
Unlike the scattering amplitude case, $\check{j}(x)$ does not receive any loop correction by definition \eqref{source_correlator}, and $\check{j}^{\ord{n}}_{x} = 0$ for $n>0$. Thus the initial condition can be computed from the recursion relation
\begin{equation}
\begin{aligned}
  \check{\Phi}^{\ord{1}}_{i} &= 
  -\frac{\lambda}{2} \frac{1}{k_{i}^{2}+m^{2}} \int_{p}
  			\check{\Phi}^{\ord{0}}_{i} \check{\Psi}^{\ord{0}}_{p|\emptyset} \,,
  \\
  &= -\frac{\hbar}{i}\frac{\lambda}{2} \frac{1}{\big(k_{i}^{2}+m^{2}\big)^{2}} \int_{p} \frac{1}{p^{2}+m^{2}} \,.
\end{aligned}\label{}
\end{equation}
This is the one-loop two-point function in momentum space. 
\paragraph{$\bullet$ Rank-2}
~\\
The one-loop rank-2 current can be determined by solving the recursion relation
\begin{equation}
\begin{aligned}
    \check{\Phi}^{\ord{1}}_{ij} &= 
  	-\frac{\lambda}{2} \frac{1}{k_{ij}^{2}+m^{2}} \bigg( \int_{p}
  			\check{\Phi}^{\ord{0}}_{i} \check{\Psi}^{\ord{0}}_{p|j} + (i\leftrightarrow j)+\int_{p}
  			\check{\Phi}^{\ord{0}}_{ij} \check{\Psi}^{\ord{0}}_{p|\emptyset} 
  		\bigg)\,.
\end{aligned}\label{}
\end{equation}
Since $\check{\Psi}^{\ord{0}}_{p|i} =0$ and $\check{\Phi}^{\ord{0}}_{ij}=0$, the rank-1 current is trivial, it is straightforward to show that $\check{\Phi}^{\ord{1}}_{ij} = 0$.
\paragraph{$\bullet$ Rank-3}
~\\
The one-loop rank-3 current satisfies 
\begin{equation}
\begin{aligned}
  \check{\Phi}^{\ord{1}}_{ijk} &= 
  	-\frac{\lambda}{2} \frac{1}{k_{ijk}^{2}+m^{2}} 
  	\sum_{\text{Perm}[i,j,k]}
  		\bigg(
  			\check{\Phi}^{\ord{0}}_{i} \check{\Phi}^{\ord{0}}_{j} \check{\Phi}^{\ord{1}}_{k} 
  			+ \frac{1}{2} \int_{p}
  			\check{\Phi}^{\ord{0}}_{i} \check{\Psi}^{\ord{0}}_{p|jk}
  			+ \frac{1}{3!} \int_{p} \check{\Phi}^{\ord{0}}_{ijk} \Psi^{\ord{0}}_{p|\emptyset}
  		\bigg)\,.
\end{aligned}\label{}
\end{equation}
Again it reproduces the 1-loop four-point function, and each term corresponds to Feynman diagrams denoted in Figure \ref{sclar_1loop_rank3_correlation}.
%

\subsection{Two-loop level}
We now consider the two-loop off-shell recursion relation. As before, the recursion relations are identical to those for computing two-loop scattering amplitudes \eqref{scalar_two_loop_recursion2}:
\begin{equation}
\begin{aligned}
  \check{\Phi}^{\ord{2}}_{\mathcal{P}} 
  &= -\frac{\lambda}{2} \frac{1}{k_{\mathcal{P}}^{2}+m^{2}} 
  	\bigg(
  		\sum_{\mathcal{P}=\mathcal{Q}\cup\mathcal{R}\cup\mathcal{S}} 
  			\Big(
  			  \check{\Phi}^{\ord{0}}_{\mathcal{Q}} \check{\Phi}^{\ord{0}}_{\mathcal{R}} \check{\Phi}^{\ord{2}}_{\mathcal{S}} 
  				+2 \check{\Phi}^{\ord{0}}_{\mathcal{Q}} \check{\Phi}^{\ord{1}}_{\mathcal{R}} \check{\Phi}^{\ord{1}}_{\mathcal{S}}
  			\Big)
  \\&\qquad\qquad\qquad\quad
  		+ \sum_{\mathcal{P}=\mathcal{Q}\cup\mathcal{R}} \int_{p} 
  			\Big(
  			  \check{\Phi}^{\ord{0}}_{\mathcal{Q}} \check{\Psi}^{\ord{1}}_{p|\mathcal{R}}
  			+ \check{\Phi}^{\ord{1}}_{\mathcal{Q}} \check{\Psi}^{\ord{0}}_{p|\mathcal{R}}
  			\Big)
  		+ \frac{1}{3} \int_{p,q} \check{\Psi}'^{\ord{0}}_{p,q|\mathcal{P}}
  	\bigg) \quad \text{for}~|\mathcal{P}|>1\,,
\end{aligned}\label{scalar_two_loop_recursion1p}
\end{equation}
which the descendant currents satisfy
\begin{equation}
\begin{aligned}
  \check{\Psi}^{\ord{1}}_{p|\mathcal{P}} &= 
  	-\frac{\lambda}{2} \frac{1}{(p-k_{\mathcal{P}})^{2}+m^{2}} 
  	\sum_{\mathcal{P}=\mathcal{Q}\cup\mathcal{R}\cup\mathcal{S}}
  		\Big( 
  			 2\check{\Phi}^{\ord{0}}_{\mathcal{Q}} \check{\Phi}^{\ord{1}}_{\mathcal{R}} \check{\Psi}^{\ord{0}}_{p|\mathcal{S}} 
  			+ \check{\Phi}^{\ord{0}}_{\mathcal{Q}} \check{\Phi}^{\ord{0}}_{\mathcal{R}} \check{\Psi}^{\ord{1}}_{p|\mathcal{S}}
  		\Big)
  	\\&\quad
  	- \frac{\lambda}{2} \frac{1}{(p-k_{\mathcal{P}})^{2}+m^{2}} 
 	  \sum_{\mathcal{P}=\mathcal{Q}\cup\mathcal{R}}\int_{q} 
  		\Big(
  		  \check{\Phi}^{\ord{0}}_{\mathcal{Q}} \check{\Psi}'^{\ord{0}}_{p,q|\mathcal{R}}
  		+ \check{\Psi}^{\ord{0}}_{p|\mathcal{Q}} \check{\Psi}^{\ord{0}}_{q|\mathcal{R}}
  		\Big)\,,
  \\
  \check{\Psi}'^{\ord{0}}_{p,q|\mathcal{P}} &= -\frac{\lambda}{2} \frac{1}{(p+q-k_{\mathcal{P}})^{2}+m^{2}}
  	\sum_{\mathcal{P}=\mathcal{Q}\cup\mathcal{R}\cup\mathcal{S}}\Big(
  		 2\check{\Phi}^{\ord{0}}_{\mathcal{Q}} \check{\Psi}_{p|\mathcal{R}}^{\ord{0}} \check{\Psi}^{\ord{0}}_{q|\mathcal{S}} 
  		+ \check{\Phi}_{\mathcal{Q}}^{\ord{0}} \check{\Phi}_{\mathcal{R}}^{\ord{0}} \check{\Psi}'^{\ord{0}}_{p,q|\mathcal{S}}
  	\Big) \,.
\end{aligned}\label{scalar_two_loop_recursion3p}
\end{equation}

Again the only difference with the recursion relation for two-loop scattering amplitudes is the initial condition. We can determine $\check{\Phi}^{\ord{2}}_{i}$ by solving \eqref{scalar_two_loop_recursion1p}\,,
\begin{equation}
  \check{\Phi}^{\ord{2}}_{i} = \frac{\lambda}{2} \frac{1}{k_{i}^{2}+m^{2} }
  	\bigg(
  	i \int_{p} 
  		\Big(
  		  \check{\Phi}^{\ord{0}}_{i} \check{\Psi}^{\ord{1}}_{p|\emptyset}
  		+ \check{\Phi}^{\ord{1}}_{i} \check{\Psi}^{\ord{0}}_{p|\emptyset}
  		\Big)
  	+ \frac{1}{3} \int_{p,q} \check{\Psi}'^{\ord{0}}_{p,q|i}
  	\bigg)\,,
\end{equation}
where the descendant currents $\check{\Psi}^{\ord{1}}_{p|\emptyset}$ and $\check{\Psi}'^{\ord{0}}_{p,q|i}$ are determined by \eqref{scalar_two_loop_recursion3p}. The first descendant field $\Psi^{\ord{1}}_{p|i}$ is given by
\begin{equation}
\begin{aligned}
  \Psi^{\ord{1}}_{p|\emptyset} 
  	&= i\frac{\lambda}{2} \frac{1}{p^{2}+m^{2}} 
  		\int_{q} \Psi^{\ord{0}}_{p|\emptyset} \Psi^{\ord{0}}_{q|\emptyset}
  \\
  	&= i\frac{\lambda}{2} \frac{1}{(p^{2}+m^{2})^{2}} 
  		\int_{q} \frac{1}{q^{2}+m^{2}} \,,
\end{aligned}\label{}
\end{equation}
and the second rank-1 descendant field $\Psi^{\ord{1}}_{p|\emptyset}$ is given by
\begin{equation}
\begin{aligned}
  \Psi'^{\ord{0}}_{p,q|i} 
  &= -\lambda \frac{1}{(p+q-k_{i})^{2}+m^{2}}
  	\Phi^{\ord{0}}_{i} \Psi_{p|\emptyset}^{\ord{0}} \Psi^{\ord{0}}_{q|\emptyset}
  \\
  &= - \frac{\lambda}{(p+q-k_{i})^{2}+m^{2}} \frac{1}{p^{2}+m^{2}} \frac{1}{q^{2}+m^{2}}\,.
\end{aligned}\label{}
\end{equation}
Collecting the necessary information, we may write down the two-loop rank-1 current
\begin{equation}
\begin{aligned}
  \check{\Phi}^{\ord{2}}_{i} = -\frac{\lambda^{2}}{k_{i}^{2}+m^{2}} \int_{p,q} 
  	\bigg(&
  		  \frac{1}{4}\frac{1}{(p^{2}+m^{2})^{2}} \frac{1}{q^{2}+m^{2}} 
  		+ \frac{1}{4}\frac{1}{p^{2}+m^{2}} \frac{1}{k_{i}^{2}+m^{2}}\frac{1}{q^{2}+m^{2}}
  	\\&
  	+ \frac{1}{6} \frac{\lambda}{(p+q-k_{i})^{2}+m^{2}} \frac{1}{p^{2}+m^{2}} \frac{1}{q^{2}+m^{2}}
  	\bigg)
\end{aligned}\label{}
\end{equation}
%

\section{Quantum Perturbiner Method for Yang--Mills theory}\label{Sec:5}
In this section, we construct the quantum perturbiner method for pure Yang--Mills (YM) theory by extending the $\phi^{4}$-theory results. Most of the structure is almost parallel with the previous case. However, there is a crucial difference: the gauge symmetry. We investigate how to treat the gauge symmetry in the framework of the quantum perturbiner method. We follow the conventional Faddeev--Popov (FP) ghost method in QFT. We impose a gauge choice and introduce ghosts to remove the unphysical gauge redundancies. We also present the perturbiner expansion for the entire ghost sector including descendants of the ghosts.

We first review the quantum effective action formalism for  pure YM theory. Then we construct the quantum perturbiner expansion by choosing the external sources appropriately. Finally, we construct the DS equation and derive the quantum off-shell recursion relation by substituting the quantum perturbiner expansion.

\subsection{Quantum effective action for pure Yang--Mills theory}
We now review the quantum effective action for the pure YM theory and establish our convention. As in the $\phi^{4}$-theory, we work with bare quantities. We choose the Feynman-'t Hooft gauge condition, $\partial_{\mu}A^{a\mu}=0$, and introduce the associated FP ghosts. The gauge fixed action with external sources for the gauge field and ghosts is given by
\begin{equation}
  S[A,c,\bar{c},j,\eta,\bar{\eta}] = \int_{x} 
  	\bigg[ 
  		- \frac{1}{4} F^{a \mu \nu} F_{\mu \nu}^{a} 
  		- \frac{1}{2}\partial^{\mu} A_{\mu}^{a} \partial^{\nu} A_{\nu}^{a} 
  		- \partial^{\mu} \bar{c}^{a} D_{\mu}^{ab} c^{b} 
  		+ A^{a}_{\mu} j^{a\mu}
 		+ \bar{c}^{a}\eta^{a}
  		+ \bar{\eta}^{a} c^{a}
  	\ \bigg]\,,
\label{}\end{equation}
where $j^{a\mu}$ is the external source for the gauge field $A^{a}_{\mu}$, $c^{a}$ and $\bar{c}^{a}$ are the FP ghosts, and $\bar{\eta}$ and $\eta^{a}$ are their external sources, respectively. Our convention for the field strength is $F_{\mu\nu}^{a} = \partial_{\mu} A_{\nu}^{a} - \partial_{\nu} A_{\mu}^{a} + g f^{abc} A_{\mu}^{b} A_{\nu}^{c}$. 

We may rewrite the action in terms of the kinetic operator and the interaction vertices,
\begin{equation}
\begin{aligned}
  S[A,c,\bar{c},j,\eta,\bar{\eta}] &= 
  	- \frac{1}{2}\int_{x,y} A^{a\mu}_{x} K^{a,b}_{\mu,\nu}(x,y) A^{b\nu}_{y} 
  	+ \frac{1}{3!}\int_{x,y,z} V^{a,b,c}_{\mu,\nu,\rho}(x,y,z) A^{a\mu}_{x} A^{b\nu}_{y} A^{b\rho}_{z}
  	\\&\quad
  	+ \frac{1}{4!} \int_{x,y,z,w} V^{a,b,c,d}_{\mu,\nu,\rho,\sigma}(x,y,z,w) A^{a\mu}_{x} A^{b\nu}_{y} A^{c\rho}_{z} A^{d\sigma}_{w}
  	- \int_{x,y} \bar{c}^{a}_{x} K^{a,b}(x,y) c^{b}_{y}
  	\\&\quad
  	+ \int_{x,y,z} V^{a,b,c}_{\mu}(x,y,z) A^{a\mu}_{x} \bar{c}^{b}_{y} c^{c}_{z}
  	+ \int_{x} \Big(
  		  A^{a\mu}_{x} j^{a\mu}_{x}
		+ \bar{c}^{a}_{x} \eta^{a}_{x}
		+ \bar{\eta}^{a}_{x} c^{a}_{x}
		\Big)\,,
\end{aligned}\label{}
\end{equation}
where $K^{a,b}_{\mu,\nu}(x,y)$ and $K^{a,b}(x,y)$ are bare kinetic operators for gluons and ghosts, respectively,
\begin{equation}
\begin{aligned}
  K^{a,b}_{\mu,\nu}(x,y) &= -\frac{\delta^{2} S}{\delta A^{a\mu}_{x} \delta A^{b\nu}_{y}}\bigg|_{A=0} = -\eta^{\mu\nu} \delta^{ab} \Box_{y} \delta^{4}(x-y) \,,
  \\
  K^{a,b}(x,y) &= - \frac{\delta^{2} S}{\delta_{L} \bar{c}^{a}_{x} \delta_{R} c^{b\nu}_{y}}\bigg|_{c=\bar{c}=0} = -\Box_{y} \delta^{4}(x-y)\,,
\end{aligned}\label{}
\end{equation}
and the interaction vertices are
\begin{equation}
\begin{aligned}
  V^{a,b,c}_{\mu,\nu,\rho}(x,y,z) &= -gf^{abc} 
  	\Big[
 		  \eta^{\mu\rho} \hat{\delta}_{xz} \partial^{z}_{\nu} \hat{\delta}_{yz}
  		- \eta^{\nu\rho} \partial^{z}_{\mu} \hat{\delta}_{xz} \hat{\delta}_{yz}
  		+ 2\eta^{\mu\rho} \partial^{z}_{\nu} \hat{\delta}_{xz} \hat{\delta}_{yz}
  		- 2\eta^{\nu\rho} \hat{\delta}_{xz} \partial^{z}_{\mu} \hat{\delta}_{yz}
  		\\&\qquad\qquad\quad
  		+ \eta^{\mu\nu} \hat{\delta}_{xz} \partial^{z}_{\rho}\hat{\delta}_{yz}
  		- \eta^{\mu\nu} \partial^{z}_{\rho} \hat{\delta}_{xz} \hat{\delta}_{yz}
  	\Big]\,,
  \\
  V^{a,b,c,d}_{\mu,\nu,\rho,\sigma}(x,y,z,w) &= -g^{2} 
  	\Big[
  		  f^{2}_{ab,cd} \big(\eta^{\mu\rho}\eta^{\nu\sigma} - \eta^{\mu\sigma}\eta^{\nu\rho}\big)
  		+ f^{2}_{ac,bd} \big(\eta^{\mu\nu}\eta^{\rho\sigma} - \eta^{\mu\sigma}\eta^{\nu\rho}\big)
  		\\&\qquad\quad
  		+ f^{2}_{ad,bc} \big(\eta^{\mu\nu}\eta^{\rho\sigma} - \eta^{\mu\rho}\eta^{\nu\sigma}\big)
  	\Big]\,,
  \\
  V^{a,b,c}_{\mu}(x,y,z) &= -f^{abc} \partial^{y}_{\mu}\hat{\delta}_{xy} \hat{\delta}_{xz}\,,
\end{aligned}\label{YM_vertices}
\end{equation}
where $\hat{\delta}_{xy} = \delta^{4}(x-y)$ and $f^{2}_{ab,cd} = f^{eab}f^{ecd}$. 

The free propagators for the gauge field and the ghosts, $D^{a,b}_{\mu,\nu}(x,y)$ and $D^{a,b}(x,y)$,respectively, are given by the inverse of the kinetic operators
\begin{equation}
\begin{aligned}
  \int_{y} K^{a,b}_{\mu,\nu}(x,y) D^{b,c}_{\nu,\rho}(y,z) &= \hat{\delta}_{xz} \delta^{ac} \eta^{\mu\rho}\,, &
  	\quad 
  D^{a,b}_{\mu,\nu}(x,y) &= \int_{p}\frac{e^{ip\cdot(x-y)}}{p^{2}} \delta^{ab} \eta^{\mu\nu}\,,
  \\
  \int_{y} K^{a,b}(x,y) D^{b,c}(y,z) &= \hat{\delta}_{xz} \delta^{ac}\,, &
  	\quad 
  D^{a,b}(x,y) &= \int_{p}\frac{e^{ip\cdot(x-y)}}{p^{2}} \delta^{ab} \,.
\end{aligned}\label{}
\end{equation}
We may denote the propagators in terms of the massless scalar propagator $D^{\ord{0}}(x-y) = D_{xy}$
\begin{equation}
  D^{a\mu,b\nu}_{xy} = D_{xy} \delta^{ab} \eta^{\mu\nu}\,, 
  \qquad 
  D^{a,b}_{xy} = D_{xy} \delta^{ab} \,,
\label{}\end{equation}
where
\begin{equation}
  D_{xy} = \int_{p} \frac{e^{ip\cdot(x-y)}}{p^{2}}\,.
\label{}\end{equation}

Let us denote the dressed propagators for the gluon and ghosts as $\mathbf{D}^{a,b}_{\mu,\nu}(x,y)$ and $\mathbf{D}^{a,b}(x,y)$, which are related to the two-point functions as
\begin{equation}
\begin{aligned}
  \mathbf{D}^{a,b}_{\mu,\nu}(x,y)&= \int_{p} e^{ip\cdot x} \tilde{\mathbf{D}}^{a,b}_{\mu,\nu}(p) = \frac{i}{\hbar} \left\langle 0|T A^{a}_{\mu}(x)A^{b}_{\nu}(y)|0\right\rangle \,, 
  \\
  \mathbf{D}^{a,b}(x,y)&= \int_{p} e^{ip\cdot x} \tilde{\mathbf{D}}^{a,b}(p) = \frac{i}{\hbar} \left\langle 0|T c^{a}(x)\bar{c}^{b}(y)|0\right\rangle \,, 
\end{aligned}\label{}
\end{equation}
At leading order, the dressed propagators reduce to the free propagators $D^{a\mu,b\nu}(x,y)$ and $D^{a,b}(x,y)$. 

The connected diagram generating functional for the pure Yang--Mills theory is defined by the functional integral 
\begin{equation}
\begin{aligned}
  e^{\frac{i}{\hbar}W[j,\eta,\bar{\eta}]} &= 
  	\int \mathcal{D}A^{a\mu}_{x} \mathcal{D}c^{a}_{x} \mathcal{D}\bar{c}^{a}_{x} \exp\bigg[
  		\frac{i}{\hbar} S[A,c,\bar{c},j,\eta,\bar{\eta}]
  		\bigg] \,.
\end{aligned}\label{WJ_gluon}
\end{equation}
We introduce classical fields $\mathcal{A}_{j}^{a\mu}(x)$, $\mathcal{C}^{a}_{j}(x)$ and $\bar{\mathcal{C}}^{a}_{j}(x)$ which are the VEVs of $A^{a}_{\mu}$, $c^{a}$ and $\bar{c}^{a}$ in the presence of their external sources $j^{a\mu}_{x}$, $\eta^{a}_{x}$ and $\bar{\eta}^{a}_{x}$. These are represented by functional derivatives of $W[j,\eta,\bar{\eta}]$,
\begin{equation}
\begin{aligned}
  \mathcal{A}^{a\mu}_{x} = \frac{\delta W[j,\eta,\bar{\eta}]}{\delta j^{a\mu}_{x}} \,, 
  \qquad 
  \mathcal{C}^{a}_{x} = \frac{\delta_{L} W[j,\eta,\bar{\eta}]}{\delta \bar{\eta}^{a}_{x}}\,,
  \qquad 
  \bar{\mathcal{C}}^{a}_{x} = \frac{\delta_{R} W[j,\eta,\bar{\eta}]}{\delta \eta^{a}_{x}}\,,
\end{aligned}\label{}
\end{equation}
where $\frac{\delta_{L}}{\delta \bar{\eta}^{a}_{x}}$ and $\frac{\delta_{R}}{\delta \eta\ind{a}{x}}$ are the left and right derivatives with respect to the Grassmannian sources. 

From the relation between functional derivatives of $W[j,\eta,\bar{\eta}]$ and connected correlation functions, we may represent the classical fields $\mathcal{A}^{a\mu}_{x}$ as
\begin{equation}
\begin{aligned}
  \mathcal{A}^{a\mu}_{x} &= \sum_{n=1}^{\infty} \frac{1}{n!} \Big(\frac{i}{\hbar}\Big)^{n}\int_{y_{1},y_{2},\cdots ,y_{n}} 
  		G_{c}{}^{a,b_{1},b_{2},\cdots,b_{n}}_{\mu,\nu_{1},\nu_{2},\cdots, \nu_{n}}(x,y_{1},y_{2}\cdots,y_{n})
  		j^{b_{1}\nu_{1}}_{y_{1}} j^{b_{2}\nu_{2}}_{y_{2}} \cdots j^{b_{n}\nu_{n}}_{y_{n}} \,,
  \\
\end{aligned}\label{classical_gluon_field}
\end{equation}
where $G_{c}{}^{a,b_{1},b_{2},\cdots,b_{n}}_{\mu,\nu_{1},\nu_{2},\cdots, \nu_{n}}(x,y_{1},y_{2}\cdots,y_{n})$ is the connected gluon correlation function
\begin{equation}
\begin{aligned}
  G_{c}{}^{a,b_{1},b_{2},\cdots,b_{n}}_{\mu,\nu_{1},\nu_{2},\cdots, \nu_{n}}(x,y_{1},y_{2}\cdots,y_{n})
  	= \left\langle0\left|T A^{a\mu}_{x} A^{b_{1}\nu_{1}}_{y_{1}} A^{b_{2}\nu_{2}}_{y_{2}} \cdots A^{b_{n}\nu_{n}}_{y_{n}}\right|0\right\rangle_{c}\,.
\end{aligned}\label{}
\end{equation}
From now on, we will ignore the position of the indices, while nevertheless keeping the summation convention for repeated indices.

We now introduce descendant fields for the gluons and ghosts that arise in the DS equation later. These are defined by acting with multiple functional derivatives on $W[j,\eta,\bar{\eta}]$,
\begin{equation}
\begin{aligned}
  &\psi\ind{a\mu,b\nu}{x,y}\,,
  \qquad
  \psi'\ind{a\mu,b\nu,c\rho}{x,y,z}\,,
  \\
  &\gamma^{a,b}_{x,y}\,,
  \\
  &\theta^{a,b\mu}_{x,y}\,,
  \qquad
  \bar{\theta}^{a,b\mu}_{x,y}\,,
  \qquad
  \text{etc.}\,,
\end{aligned}\label{}
\end{equation}
where
\begin{equation}
\begin{aligned}
    \psi\ind{a\mu,b\nu}{x,y} &= \frac{\delta A\ind{a\mu}{x}}{\delta j\ind{b\nu}{y}} =  \frac{\delta^{2} W[j]}{\delta j\ind{a\mu}{x}\delta j\ind{b\nu}{y}} \,, 
  \\
  \psi'\ind{a\mu,b\nu,c\rho}{x,y,z} &= \frac{\delta^{2} A\ind{a\mu}{x}}{\delta j\ind{b\nu}{y} \delta j\ind{c\rho}{z}} =\frac{\delta^{3} W[j]}{\delta j\ind{a\mu}{x} \delta j\ind{b\nu}{y} \delta j\ind{c\rho}{z}} \,,
  \\
  \gamma^{a,b}_{x,y} &= \frac{\delta_{L}c^{a}_{x}}{\delta \eta^{b}_{x}} 
    = \frac{\delta_{R}c^{a}_{x}}{\delta \bar{\eta}^{b}_{x}} 
    = \frac{\delta^{2} W[J,\eta,\bar{\eta}]}{\delta \eta^{a}_{x} \delta \bar{\eta}^{b}_{x}}\,,
    \\
    \theta^{a,b\mu}_{x,y} &= \frac{\delta c^{a}_{x}}{\delta j^{b\mu}_{y}}
    = \frac{\delta^{2} W[J,\eta,\bar{\eta}]}{\delta \eta^{a}_{x} \delta j^{b\mu}_{y}}\,,
    \\
    \bar{\theta}^{a,b\mu}_{x,y} &= \frac{\delta \bar{c}^{a}_{x}}{\delta j^{b\mu}_{y}}
    = \frac{\delta^{2} W[J,\eta,\bar{\eta}]}{\delta \bar{\eta}^{a}_{x} \delta j^{b\mu}_{y}}\,.
\end{aligned}\label{}
\end{equation}

Since the classical fields and correlation functions are related, we may write an $\hbar$-expansion for all the fields and descendant fields
\begin{equation}
\begin{aligned}
  \mathcal{A}^{a\mu}_{x} = \sum_{n} \mathcal{A}^{\ord{n} a\mu}_{x}\Big(\frac{\hbar}{i}\Big)^{n} \,, 
  \qquad  
  \psi^{a\mu,b\nu}_{x,y} = \sum_{n} \psi^{\ord{n} a\mu,b\nu}_{x,y}\Big(\frac{\hbar}{i}\Big)^{n} \,,
  \qquad 
  \psi'^{a\mu,b\nu,c\rho}_{x,y,z} = \sum_{n} \psi'^{\ord{n} a\mu,b\nu}_{x,y} \Big(\frac{\hbar}{i}\Big)^{n} \,.
\end{aligned}\label{}
\end{equation}
%

\subsection{Perturbiner expansion for gluon off-shell currents}
As we have shown in $\phi^{4}$-theory, we take the classical field $\mathcal{A}^{a\mu}_{x}$ as the field defining the quantum perturbiner expansion. To this end, we choose the external source $j^{a\mu}_{x}$ to reproduce the LSZ reduction formula from the expansion of $W[j,\eta,\bar{\eta}]$. For Yang--Mills theory, there are two types of perturbiner expansions, color-stripped and color-dressed perturbiners \cite{Mizera:2018jbh}. In this paper, we employ the color-dressed perturbiner, which carries the color indices explicitly,
\begin{equation}
\begin{aligned}
  \mathcal{A}^{a}_{\mu} = \sum_{\mathcal{P}} J^{a\mu}_{\mathcal{P}} e^{-ik_{\mathcal{P}}\cdot x}\,,
\end{aligned}\label{}
\end{equation}
where $\mathcal{P}$ is the set of ordered words. It leads to the Del Duca--Dixon--Maltoni half-ladder basis of partial amplitudes \cite{DelDuca:1999rs,Mizera:2018jbh}.

We now assign external sources for gluons, $j^{a\mu}_{x} :=j^{a\mu}(x)$, and ghost fields, $\eta^{a}_{x} = \eta^{a}(x)$ and $\bar{\eta}^{a}_{x}:=\bar{\eta}^{a}(x)$, as follows:
\begin{equation}
\begin{aligned}
  j^{a\mu}_{x} &= \sum_{i=1}^{N}\int_{y_{i}} \, \mathbf{K}^{a,b_{i}}_{\mu,\nu_{i}}(x, y_{i}) \epsilon^{\nu_{i}}_{i} e^{-ik_{i}\cdot y_{i}} 
  = \sum_{i=1}^{N}\, \tilde{\mathbf{K}}^{a,b_{i}}_{\mu,\nu_{i}}(-k_{i}) \epsilon^{\nu_{i}}_{i}e^{-ik_{i}\cdot x} \,,
  \\
  \eta\ind{a}{x} &= 0\,,
  \qquad\qquad
  \bar{\eta}\ind{a}{x} = 0\,,
\end{aligned}\label{gluon_source}
\end{equation}
where $k_{i}$ are on-shell momenta for external gluons and $\epsilon^{\nu_{i}}_{i}$ are their polarization vectors. Note that the external sources for the ghosts vanish because their external on-shell states do not exist. However, it does not mean that the entire ghost sector vanishes -- we cannot ignore their descendant fields such as $\gamma^{a,b}_{x,y}$ and $\theta^{a,b\mu}_{x,y}$. Similar to the $\phi^{4}$-theory in \eqref{scalar_LSZ}, we can check that this choice of $j^{a\mu}_{x}$ yields gluon scattering amplitudes from $W[j,\eta,\bar{\eta}]$ through the LSZ reduction formula. 

Since we are considering one-loop gluon scattering amplitudes, we expand $j_{x}$ up to one-loop order by evaluating the one-loop gluon self-energy including the ghost loop,
\begin{equation}
\begin{aligned}
  j^{\ord{0}a\mu}_{x} &= \sum_{i=1}^{N} k_{i}^{2} \delta^{ab_{i}} \epsilon^{\mu}_{i} e^{-ik_{i}\cdot x}\,,
  \\ 
  j^{\ord{1}a\mu}_{x} &= -\frac{T(A)}{2} \sum_{i=1}^{N} \int_{p} \frac{N_{i}^{\mu\nu_{i}}+(p^{\mu}-k_{i}^{\mu})p^{\nu_{i}}}{p^{2}(p-k_{i})^{2}} \delta^{ab_{i}}\epsilon^{\nu_{i}}_{i} e^{-ik_{i}\cdot x}\,,
\end{aligned}\label{}
\end{equation}
where $f^{a c d} f^{b c d}=T(\mathrm{A}) \delta^{a b}$ and 
\begin{equation}
\begin{aligned}
  N^{\mu\nu}_{i} &= 
  -\Big[(2 p -k_{i})^{\mu} \eta^{\rho \sigma}-(p+k_{i})^{\rho} \eta^{\sigma \mu}-(p-2 k_{i})^{\sigma} \eta^{\mu \rho}\Big] 
  \\&\quad 
  \times\Big[(2p -k_{i})^{\nu} \eta_{\rho \sigma}-(p +k_{i})_{\rho} \delta_{\sigma}{}^{\nu}-(p -2 k_{i})_{\sigma} \delta^{\nu}{ }_{\rho}\Big]\,.
\end{aligned}\label{}
\end{equation}

We may construct the quantum perturbiner expansion for the gluon by substituting the relation between the classical field and its external source \eqref{classical_gluon_field}, 
\begin{equation}
\begin{aligned}
  \mathcal{A}^{a\mu}_{x} &= \sum_{i} J^{a\mu}_{i} e^{-ik_{i} \cdot x} + \sum_{i,j} J^{a\mu}_{ij} e^{-ik_{ij} \cdot x} + \cdots + \sum_{i_{1},i_{2}\cdots ,i_{n}} J^{a\mu}_{i_{1}i_{2}\cdots i_{n}} e^{-ik_{i_{1}i_{2}\cdots i_{n}}}+\cdots
  \\
  & = \sum_{\mathcal{P}} J^{a\mu}_{\mathcal{P}} e^{-ik_{\mathcal{P}}\cdot x}\,,
\end{aligned}\label{gluon_perturbiner}
\end{equation}
and we identify the coefficients $J^{a\mu}_{\mathcal{P}}$ as the \emph{quantum off-shell gluon currents}. The ghost currents should be trivial because there are no on-shell external states for ghosts. Here we have adopted the color-dressed perturbiner expansion, and the color-dressed quantum off-shell currents correspond to Del Duca--Dixon--Maltoni half-ladder basis of partial amplitudes.

Substituting our external sources \eqref{gluon_source} into the classical gluon field \eqref{classical_gluon_field}, we can represent the quantum off-shell gluon currents in terms of the connected correlation functions,
\begin{equation}
\begin{aligned}
  J^{a\mu}_{i_{1}i_{2}\cdots i_{n}} &= \bigg(\frac{i}{\hbar}\bigg)^{n}
  		\tilde{G}_{c}{}^{a,b_{1},\cdots,b_{n}}_{\mu,\nu_{1},\cdots, \nu_{n}}(-k_{i_{1}i_{2}\cdots i_{n}},k_{i_{1}},\cdots,k_{i_{n}})
  		\tilde{\mathbf{K}}^{b_{1},c_{i_{1}}}_{\nu_{1},\rho_{i_{1}}}(k_{i_{1}}) \cdots \tilde{\mathbf{K}}^{b_{n},c_{i_{n}}}_{\nu_{n},\rho_{i_{n}}}(k_{i_{n}}) \epsilon^{\rho_{i_{1}}}_{i_{1}} \cdots \epsilon^{\rho_{i_{n}}}_{i_{n}}
\end{aligned}\label{}
\end{equation}
As for $\phi^{4}$-theory, we require that the off-shell currents with repeated momenta vanish, 
\begin{equation}
  J^{a\mu}_{i_{1}\cdots j\cdots j\cdots i_{n}} = 0\,,
\label{}\end{equation}
and this prohibits higher-rank currents than the number of external particles,
\begin{equation}
  J^{a\mu}_{\mathcal{P}}= 0 \qquad\text{for}~|\mathcal{P}|>N\,.
\label{}\end{equation}
We also introduce the quantum perturbiner expansion for the descendant fields,
\begin{equation}
\begin{aligned}
  \psi{}^{a\mu,b\nu}_{x,y} &= \Psi^{a\mu,b\nu}_{p|\emptyset} e^{ip\cdot(x-y)} + \sum_{\mathcal{P}} \int_{p} \Psi^{a\mu,b\nu}_{p|\mathcal{P}} e^{ip\cdot(x-y)} e^{-ik_{\mathcal{P}}\cdot x}\,, 
  \\
  \psi'{}^{a\mu,b\nu,c\rho}_{x,y,z} &= \sum_{\mathcal{P}} \int_{p,q}\Psi'^{a\mu,b\nu,c\rho}_{p,q|\mathcal{P}} e^{ip\cdot(x-y)+iq\cdot (x-z)}e^{-ik_{\mathcal{P}}\cdot x}\,, 
  \\
  \gamma{}^{a,b}_{x,y} &= \sum_{\mathcal{P}} \int_{p} \Gamma^{a,b}_{p|\mathcal{P}} e^{ip\cdot(x-y)}e^{-ik_{\mathcal{P}}\cdot x}\,, 
\end{aligned}\label{YM_descendant_perturbiner}
\end{equation}
and the higher-order descendant fields are similarly defined. Again, $p$ and $q$ are internal loop-momenta. As before, the perturbiner equation is expanded in powers of $\hbar$,
\begin{equation}
  J^{a\mu}_{\mathcal{P}} = \sum_{n=0}^{\infty} J^{(n)a\mu}_{\mathcal{P}} \Big(\frac{\hbar}{i}\Big)^{n}\,, 
  \qquad \mathcal{C}^{a}_{x} = \sum_{n=0}^{\infty}\mathcal{C}^{(n)a}_{\mathcal{P}} \Big(\frac{\hbar}{i}\Big)^{n}\,, 
  \qquad \bar{\mathcal{C}}^{a}_{x} = \sum_{n=0}^{\infty}\bar{\mathcal{C}}^{(n)a}_{\mathcal{P}} \Big(\frac{\hbar}{i}\Big)^{n}\,.
\label{hbar_expan_current}\end{equation}

According to the LSZ reduction formula, scattering amplitudes arise from the amputation of off-shell currents. The $n$-loop $(N+1)$-point scattering amplitude is given by
\begin{equation}
\begin{aligned} 
  \mathcal{A}^{\ord{n}}\big(k_{1},\cdots,k_{N+1}\big) 
  &= 
  \frac{i}{\hbar} \lim _{k_{12\cdots N}^{2} \to 0} \sum_{p+q=n} \epsilon^{\mu}_{N+1} \delta^{a,a_{N+1}} \tilde{\mathbf{K}}^{\ord{p}}{}^{a,b}_{\mu,\nu}(-k_{12\cdots N}) J^{\ord{q}}{}^{b\nu}_{1\cdots N}\,.
\end{aligned}
\label{}\end{equation}
The color decomposition in the Del Duca--Dixon--Maltoni half-ladder basis $\mathcal{F}_{1 \rho(23 \cdots N)}^{a_{N+1}}$ is given by
\begin{equation}
  \mathcal{F}_{12 \cdots m}^{a}=f_{a_{1} a_{2} b} f_{b a_{3} c} \cdots f_{d a_{m-1} e} f_{e a_{m} a} \,.
\label{}\end{equation}
We define the amputated off-shell current as
\begin{equation}
  \hat{J}^{\ord{n}}{}^{a\mu}_{\mathcal{P}} = \frac{i}{\hbar} \sum_{p+q=n} \tilde{\mathbf{K}}^{\ord{p}}{}^{a,b}_{\mu,\nu}(-k_{\mathcal{P}}) J^{\ord{q}}{}^{b\nu}_{\mathcal{P}}\,.
\label{}\end{equation}
%
\subsection{DS equation for YM theory}
As we have seen in $\phi^{4}$-theory, we can derive the DS equation for pure YM theory from the classical EoM by deforming $\mathcal{A}^{a\mu}_{x} \to \mathcal{A}^{a\mu}_{x} + \frac{\hbar}{i} \frac{\delta}{\delta j^{a\mu}_{x}}$, $\mathcal{C}^{a} \to \mathcal{C}^{a} + \frac{\hbar}{i}\frac{\delta_{L}}{\delta \bar{\eta}\ind{a}{x}}$ and $\bar{\mathcal{C}}^{a} \to \bar{\mathcal{C}}^{a} + \frac{\hbar}{i} \frac{\delta_{R}}{\delta \eta\ind{a}{x}}$. 
Substituting the deformation into the classical EoM, we have 
\begin{equation}
\begin{aligned}
  \int_{y }K_{x,y}^{a\mu,b\nu} \mathcal{A}^{b\nu}_{y} &= 
  	  j^{a\mu}_{x}
  	+ \frac{1}{2!} \int_{y,z} V{}^{a,b,c}_{\mu,\nu,\rho}(x,y,z) \Big(\mathcal{A}^{b\nu}_{y} \mathcal{A}^{c\rho}_{z} -i\hbar \psi^{c\rho,b\nu}_{z,y}\Big)
  	\\&\quad
    + \frac{1}{3!} \int_{y,z,w} V{}^{a,b,c,d}_{\mu,\nu,\rho,\sigma}(x,y,z,w) 
  		\bigg[\
   			\mathcal{A}^{b\nu}_{y} \mathcal{A}^{c\rho}_{z} \mathcal{A}^{d\sigma}_{w} - \hbar^{2} \psi'{}^{d\sigma,c\rho,b\nu}_{w,z,y}
  			\\&\qquad\qquad\qquad\qquad\qquad\qquad\qquad\
   			{+}\frac{\hbar}{i} 
   			\Big(
   			 	  \mathcal{A}^{b\nu}_{y}\psi^{d\sigma,c\rho}_{w,z}
  				{+} \mathcal{A}^{c\rho}_{z}\psi^{d\sigma,b\nu}_{w,y}
  				{+} \mathcal{A}^{d\sigma}_{w}\psi^{c\rho,b\nu}_{z,y}
  			\Big)
  		\bigg]
  	\\&\quad
  	+ \int_{y,z} V^{a,b,c}_{\mu}(x,y,z) \Big(\bar{\mathcal{C}}^{b}_{y} \mathcal{C}^{c}_{z} - i\hbar \gamma^{c,b}_{z,y} \Big)\,,
  \\
  \int_{y} K^{a,b}_{x,y} \mathcal{C}^{b}_{y} &=
  	  \eta^{a}_{x}
  	+ \int_{y,z} V^{a,b,c}_{\mu}(x,y,z) \Big(\mathcal{A}^{a\mu}_{y} \mathcal{C}^{c}_{z} -i\hbar \theta^{c,a\mu}_{z,y} \Big) \,,
  \\
  \int_{y} K^{a,b}_{x,y} \bar{\mathcal{C}}^{b}_{y} &=
  	  \bar{\eta}^{a}_{x}
  	+ \int_{y,z} V^{b,c,a}_{\mu}(x,y,z) \Big(\mathcal{A}^{b\mu}_{y} \bar{\mathcal{C}}^{c}_{z} -i\hbar \bar{\theta}^{c,a\mu}_{z,y} \Big) \,,
\end{aligned}\label{YM_DSeq1}
\end{equation}
If we evaluate \eqref{YM_DSeq1} by using the interaction vertices \eqref{YM_vertices}, we get
\begin{equation}
\begin{aligned}
  \int_{y} K_{x,y} \mathcal{A}_{y}^{a\mu} &= 
  	j^{a\mu}_{x} 
  	+ gf_{abc} 
  		\Big( 
  		    \mathcal{A}^{b\nu}_{x} \big(2\partial_{\nu} \mathcal{A}^{c\mu}_{x} -\partial_{\mu} \mathcal{A}^{c\nu}_{x}\big) 
  		  - \mathcal{A}^{b\mu}_{x} \partial_{\nu} \mathcal{A}^{c\nu}_{x}
  		\Big)
  	- g^{2} f^{2}_{ab,cd} \mathcal{A}^{b\nu}_{x} \mathcal{A}^{c\mu}_{x} \mathcal{A}^{d\nu}_{x} 
    \\&\quad 
  	+ \frac{i\hbar}{2} g f^{abc} \lim_{y\to x}
  		\Big[
  			~ \big(2\partial^{y}_{\nu} +\partial^{x}_{\nu} \big) \big(\psi{}^{c\nu,b\mu}_{x,y} 
  			+ \psi{}^{b\mu,c\nu}_{y,x}\big)
  			+ \partial^{x}_{\mu} \big(\psi{}^{c\nu,b\nu}_{x,y} + \psi{}^{b\nu,c\nu}_{y,x}\big)
  		\Big]
    \\&\quad 
  	+ i\hbar g^{2} f^{2}_{ab,cd} \Big( 
  	    \mathcal{A}^{b\nu}_{x} \psi^{\{d\nu,c\mu\}}_{x,x} 
  	  + \mathcal{A}^{c\mu}_{x} \psi^{\{d\nu,b\nu\}}_{x,x} 
  	  + \mathcal{A}^{d\nu}_{x} \psi^{\{b\nu,c\mu\}}_{x,x} 
  	\Big) 
  	\\&\quad 
  	+ \hbar^{2} g^{2} f^{2}_{ab,cd} \psi'{}^{b\nu,c\mu,d\nu}_{x,x,x}
  	+ gf^{abc} 
  		\Big(
  			  \partial_{\mu}\bar{\mathcal{C}}^{b}_{x} c^{c}_{x}
  			- i\hbar \lim_{y\to x}\partial^{y}_{\mu} \gamma^{c,b}_{x,y}
  		\Big) \,,
  \\
  \int_{y}K^{ab}_{xy} \mathcal{C}_{y}^{b} &= 
  	  \eta^{a}_{x} 
  	+ g f^{abc} 
  		\bigg(
  			  \partial_{\mu} \mathcal{A}^{b\mu}_{x} \mathcal{C}^{c}_{x} 
  			+ \mathcal{A}^{b\mu}_{x} \partial_{\mu} \mathcal{C}^{c}_{x} 
  			- i\hbar \, \partial^{x}_{\mu} \theta^{c,b\mu}_{x,x} 
  		\bigg)\,,
    \\
  \int_{y}K^{ab}_{xy} \bar{\mathcal{C}}_{y}^{b} &=  
  	  \bar{\eta}^{a}_{x}
  	+ gf^{abc} \bigg(\mathcal{A}^{b\mu}_{x}\partial_{\mu} \bar{\mathcal{C}}^{c}_{x}  -i\hbar \, \partial^{z}_{\mu} \theta^{c,b\mu}_{z,x} \bigg) \,.
\end{aligned}\label{gluon_DSe}
\end{equation}
where the curly bracket $\{a,b,c\cdots\}$ in the third line denotes symmetrisation over the pair of indices $a\mu$ and $b\nu$,
\begin{equation}
  \psi{}^{\{d\nu,c\mu\}}_{x,x} = \frac{1}{2} \Big(\psi^{d\nu,c\mu}_{x,x} + \psi^{c\mu,d\nu}_{x,x}\Big)\,.
\label{}\end{equation}

We construct the field equations for the descendant fields $\psi^{a\mu,b\nu}_{x,y}$, $\gamma^{a,b}_{x,y}$ and $\bar{\gamma}^{a,b}_{x,y}$ by acting with functional derivatives with respect to the external sources, $\frac{\delta}{\delta j^{a\mu}_{x}}$, $\frac{\delta}{\delta\eta^{a}_{x}}$ and $\frac{\delta}{\delta\bar{\eta}^{a}_{x}}$, respectively, on both sides of the DS equation: for $\psi^{a\mu,b\nu}_{x,y}$,
\begin{equation}
\begin{aligned}
  \int_{y}K_{xy} \psi_{y,z}^{a\mu,e\rho} &=
    \delta^{ae} \eta^{\mu\rho}\hat{\delta}_{xz}
  + gf^{abc} \Big( 
  	  \psi^{b\nu,e\rho}_{x,z} \big(
  	  	2\partial_{\nu} \mathcal{A}^{c\mu}_{x} 
  	  	-\partial_{\mu} \mathcal{A}^{c\nu}_{x}
  	  	\big) 
  	+ \mathcal{A}^{b\nu}_{x} \big(
  		2 \partial^{x}_{\nu} \psi^{c\mu,e\rho}_{x,z} 
  		- \partial^{x}_{\mu} \psi_{x,z}^{c\nu,e\rho}
  		\big)
  	\Big)
  \\&\quad 
  - gf^{abc} \Big(
  	  \psi^{b\mu,e\rho}_{x,z} \partial_{\nu} \mathcal{A}^{c\nu}_{x} 
  	+ \mathcal{A}^{b\mu}_{x} \partial^{x}_{\nu} \psi^{c\nu,e\rho}_{x,z}
  \Big) 
  \\&\quad 
  - g^{2} f^{2}_{ab,cd} \Big(
      \psi^{b\nu,e\rho}_{x,z} \mathcal{A}^{c\mu}_{x} \mathcal{A}^{d\nu}_{x}
  	+ \mathcal{A}^{b\nu}_{x} \psi_{x,z}^{c\mu,e\rho} \mathcal{A}^{d\nu}_{x} 
  	+ \mathcal{A}^{b\nu}_{x} \mathcal{A}_{x}^{c\mu} \psi^{d\nu,e\rho}_{x,z}
  \Big)
  \\&\quad 
  + \frac{i\hbar}{2} g f^{abc} \lim_{y\to x}
  	\Big[
  		  \big(2\partial^{y}_{\nu} +\partial^{x}_{\nu} \big) \big(\psi{}^{c\nu,b\mu,e\rho}_{x,y,z} 
  		{+} \psi{}^{b\mu,c\nu,e\rho}_{y,x,z}\big)
  		+ \partial^{x}_{\mu} \big(\psi{}^{c\nu,b\nu,e\rho}_{x,y} {+} \psi{}^{b\nu,c\nu,e\rho}_{y,x,z}\big)
  	\Big]
  \\&\quad 
  + i\hbar g^{2} f^{2}_{ab,cd} \Big( 
  	    A^{b\nu}_{x} \psi^{\{d\nu,c\mu\},e\rho}_{x,x,z} 
  	  + A^{c\mu}_{x} \psi^{\{d\nu,b\nu\},e\rho}_{x,x,z} 
  	  + A^{d\nu}_{x} \psi^{\{b\nu,c\mu\},e\rho}_{x,x,z} 
  	  \\&\qquad\qquad\qquad~
  	  + \psi^{b\nu,e\rho}_{x,z} \psi^{\{d\nu,c\mu\}}_{x,x} 
  	  + \psi^{c\mu,e\rho}_{x,z} \psi^{\{d\nu,b\nu\}}_{x,x} 
  	  + \psi^{d\nu,e\rho}_{x,z} \psi^{\{b\nu,c\mu\}}_{x,x} 
  	\Big)
  \\&\quad 
  + \hbar^{2} g^{2} f^{2}_{ab,cd} \psi''{}^{b\nu,c\mu,d\nu,e\rho}_{x,x,x,z}   	
  + gf^{abc} \bigg(
  	  \partial^{x}_{\mu}\frac{\delta \bar{c}^{b}_{x}}{\delta j^{e\rho}_{z}} c^{c}_{x} 
  	+ \partial^{x}_{\mu}\bar{c}^{b}_{x} \frac{\delta c^{c}_{x}}{\delta j^{e\rho}_{x}}
  	\bigg) 
\end{aligned}\label{gluon_DSe1}
\end{equation}
and for $\gamma^{a,b}_{x,y}$ and $\bar{\gamma}^{a,b}_{x,y}$,
\begin{equation}
\begin{aligned}
  \int_{y}K^{ab}_{xy} \gamma_{y,z}^{b,d} &= 
    \hat{\delta}_{xz} \delta^{ad}
  +	gf^{abc} \bigg(
  		\partial^{x}_{\mu} \Big( \mathcal{A}^{b\mu}_{x} \gamma^{c,d}_{x,z} \Big)
  		+ \partial_{\mu} \Big( \theta^{b\mu,d}_{x,z} \mathcal{C}^{c}_{x} \Big) 
  		-i\hbar \, \partial^{x}_{\mu} \theta^{c,d,b\mu}_{x,x,z}
  	\bigg) \,,
  \\
  \int_{y}K^{ab}_{xy} \bar{\gamma}_{y,z}^{b,d} &= 
    \hat{\delta}_{xz} \delta^{ad}
  + gf^{abc} \bigg(
  		  \mathcal{A}^{b\mu}_{x} \partial_{\mu}^{x} \bar{\gamma}^{c,d}_{x,z}  
  		+ \theta^{b\mu,d}_{x,z} \bar{\mathcal{C}}^{c}_{x}
  		- i\hbar \lim_{y\to x} \partial^{y}_{\mu} \theta^{c,d,b\mu}_{x,y,z} 
  	\bigg) \,.
\end{aligned}\label{}
\end{equation}
We will construct the recursion relations for the off-shell current and the descendant fields using the above results.

\subsubsection{Tree level}
In tree-level case, the DS equation reduces to the classical EoM, and all the descendant fields disappear. Further, the ghosts are completely decoupled, and we can focus on the gluon currents. Therefore, the recursion relations for the gluon currents is the same as the conventional Berends-Giele (BG) recursion relations \cite{Berends:1987me} in the color-dressed version. From now on, we denote all the tree-level quantities by attaching a tilde for brevity, such as $\tilde{A}^{a\mu}_{x} = A^{\ord{0}a\mu}_{x}$, $\tilde{\psi}{}^{a\mu,b\nu}_{x,y} = \psi^{\ord{0}a\mu,b\nu}_{x,y}$, etc. The classical EoM for the gauge field in the Feynman-'t Hooft gauge is given by
\begin{equation}
\begin{aligned}
   \tilde{\mathcal{A}}^{a\mu}_{x} &= \int_{y} D_{xy} \bigg[
   	  j^{a\mu}_{y}
  	+ gf^{abc} \Big( 
  		\tilde{\mathcal{A}}^{b\nu}_{y} \big(2\partial_{\nu} \tilde{\mathcal{A}}^{c\rho}_{y} 
  		- \partial_{\rho} \tilde{A}^{c\nu}_{y}\big) 
  		- \tilde{\mathcal{A}}^{b\rho}_{y} \partial_{\nu} \tilde{\mathcal{A}}^{c\nu}_{y}
  		\Big) 
  	- g^{2} f^{2}_{ab,cd} \tilde{\mathcal{A}}^{b\nu}_{y} \tilde{\mathcal{A}}^{c\mu}_{y} \tilde{\mathcal{A}}^{d\nu}_{y}
  	\bigg]\,.
\end{aligned}\label{}
\end{equation}

Substituting the perturbiner expansions defined in \eqref{gluon_perturbiner} and \eqref{YM_descendant_perturbiner} into the tree-level DS equation, we reproduce the BG recursion relation
\begin{equation}
\begin{aligned}
  \tilde{J}^{a\mu}_{\mathcal{P}} &= -\frac{1}{k_{\mathcal{P}}^{2}} 
  \bigg[\
  	igf^{abc} \sum_{\mathcal{P}=\mathcal{Q}\cup\mathcal{R}}  
  	\Big(
  		\tilde{J}^{b\nu}_{\mathcal{Q}} 
  			\big( 
  				 2k^{\nu}_{\mathcal{R}} \tilde{J}^{c\mu}_{\mathcal{R}}
  				- k^{\mu}_{\mathcal{R}} \tilde{J}^{c\nu}_{\mathcal{R}}
  			\big) 
  		- \tilde{J}^{b\mu}_{\mathcal{Q}} k^{\nu}_{\mathcal{R}} \tilde{J}^{c\nu}_{\mathcal{R}}
  	\Big)
  	\\&\qquad \qquad 
  	+ g^{2} f^{2}_{ab,cd} \sum_{\mathcal{P}=\mathcal{Q}\cup\mathcal{R}\cup\mathcal{S}}
      \tilde{J}^{b\nu}_{\mathcal{Q}} \tilde{J}^{c\mu}_{\mathcal{R}} \tilde{J}^{d\nu}_{\mathcal{S}}
  \ \bigg]\,.
\end{aligned}\label{}
\end{equation}
In our formulation, we can derive the initial condition for the gluon current $\tilde{J}^{a\mu}_{i}$. The DS equation at rank 1 is
\begin{equation}
  \sum_{i} \tilde{J}_{i}^{a\mu} e^{-ik_{i}\cdot x} = \int_{y} D_{xy} j^{a\mu}_{y} = \sum_{i}\int_{y} D_{xy} k^{2} \epsilon^{\mu}_{i} \delta^{ab_{i}}e^{-ik_{i}\cdot y}\,,
\label{}\end{equation}
and we can easily read off the rank-1 current
\begin{equation}
  \tilde{J}^{a\mu}_{i} = \epsilon^{\mu}_{i} \delta^{ab_{i}} \,.
\label{}\end{equation}
This is the same as the initial condition of the BG recursion relation.

\subsubsection{One loop}
We now consider the DS equation at the one-loop level. We denote the one-loop quantities that arise in the one-loop DS equations as follows:
\begin{equation}
\begin{aligned}
  \dot{\mathcal{A}}\ind{a\mu}{x} = \mathcal{A}\ind{\ord{1}a\mu}{x}\,, 
  \qquad 
  \tilde{\psi}\ind{a\mu,b\nu}{x,y} = \psi\ind{\ord{0}}{}\ind{a\mu,b\nu}{x,y} \,,
  \qquad
  \tilde{\gamma}\ind{a,b}{x,y} = \gamma^{\ord{0}}\ind{a,b}{x,y} \,.
\end{aligned}\label{}
\end{equation}
The perturbiner expansions of these fields are given by
\begin{equation}
\begin{aligned}
  \dot{\mathcal{A}}^{a\mu}_{x} &= \sum_{\mathcal{P}} \dot{J}^{a\mu}_{\mathcal{P}} e^{-ik_{\mathcal{P}}\cdot x}\,,
  \\
  \tilde{\psi}^{a\mu,b\nu}_{x,y} &= \int_{p} \tilde{\Psi}^{a\mu,b\nu}_{p|\emptyset} e^{ip\cdot(x-y)} + \sum_{\mathcal{P}} \int_{p} \tilde{\Psi}^{a\mu,b\nu}_{p|\mathcal{P}} e^{ip\cdot(x-y)} e^{-ik_{\mathcal{P}}\cdot x}\,,
  \\
  \tilde{\gamma}^{a,b}_{x,y} &= \sum_{\mathcal{P}} \int_{p} \tilde{\Gamma}^{a,b}_{p|\mathcal{P}} e^{ip\cdot(x-y)} e^{-ik_{\mathcal{P}}\cdot x}\,,
\end{aligned}\label{}
\end{equation}
where the off-shell currents are denoted by $\dot{J}^{a\mu}_{\mathcal{P}} = J^{\ord{1}}{}^{a\mu}_{\mathcal{P}}$, $\tilde{\Psi}^{a\mu,b\nu}_{p|\mathcal{P}} = \Psi^{\ord{0}}{}^{a\mu,b\nu}_{p|\mathcal{P}}$ and $\tilde{\Gamma}^{a,b}_{p|\mathcal{P}} = \Gamma^{\ord{0}}{}^{a,b}_{p|\mathcal{P}}$.

As before, we expand the DS equations \eqref{gluon_DSe} and \eqref{gluon_DSe1} in $\hbar$ and collect the first-order terms:
\begin{equation}
\begin{aligned}
  \dot{\mathcal{A}}^{a\mu}_{x}  &= 
  gf^{abc} \int_{y} D_{xy} \Big[
      \tilde{\mathcal{A}}^{b\nu}_{y} \big(2\partial_{\nu} \dot{\mathcal{A}}^{c\mu}_{y} {-} \partial_{\mu}\dot{\mathcal{A}}^{c\nu}_{y}\big) 
  	{-} \tilde{\mathcal{A}}^{b\mu}_{y} \partial_{\nu} \dot{\mathcal{A}}^{c\nu}_{y}
  	{+} \dot{\mathcal{A}}^{b\nu}_{y} \big(
  		2\partial_{\nu} \tilde{\mathcal{A}}^{c\mu}_{y} 
  		{-}\partial_{\mu} \tilde{\mathcal{A}}^{c\nu}_{y}
  		\big)
  	{-}\dot{\mathcal{A}}^{b\mu}_{y} \partial_{\nu} \tilde{\mathcal{A}}^{c\nu}_{y}
  	\Big]
  \\ &\quad 
  - g^{2}f^{2}_{ab,cd} \int_{y} D_{xy} \Big[\
  	  \dot{\mathcal{A}}^{b\nu}_{y} \tilde{\mathcal{A}}^{c\mu}_{y} \tilde{\mathcal{A}}^{d\nu}_{y} 
  	+ \tilde{\mathcal{A}}^{b\nu}_{y} \dot{\mathcal{A}}^{c\mu}_{y} \tilde{\mathcal{A}}^{d\nu}_{y} 
  	+ \tilde{\mathcal{A}}^{b\nu}_{y} \tilde{\mathcal{A}}^{c\mu}_{y} \dot{\mathcal{A}}^{d\nu}_{y}
  	\Big]
  \\&\quad 
  - \frac{g}{2} f^{abc}\lim_{z\to y}\int_{y}D_{xy}
  \Big[
  		~ \big(2\partial^{z}_{\nu} + \partial^{y}_{\nu} \big) \big(\tilde{\psi}{}^{c\nu,b\mu}_{y,z} 
  		+ \tilde{\psi}{}^{b\mu,c\nu}_{z,y}\big)
  		+ \partial^{y}_{\mu} \big(\tilde{\psi}{}^{c\nu,b\nu}_{y,z} + \tilde{\psi}{}^{b\nu,c\nu}_{z,y}\big)
  	\Big]
  \\&\quad 
  - g^{2} f^{2}_{ab,cd} \int_{y} D_{xy} 
  \Big[\
    	\tilde{A}^{b\nu}_{x} \tilde{\psi}^{\{d\nu,c\mu\}}_{y,y} 
  	  + \tilde{A}^{c\mu}_{x} \tilde{\psi}^{\{d\nu,b\nu\}}_{y,y} 
  	  + \tilde{A}^{d\nu}_{x} \tilde{\psi}^{\{b\nu,c\mu\}}_{y,y} 
  \Big]
  \\&\quad 
  + g f^{abc} \lim_{z\to y} \int_{y} D_{xy} 
    \partial^{y}_{\mu} \tilde{\gamma}^{c,b}_{y,z} 
  + \int_{y} D_{xy} j^{\ord{1} a\mu}_{y}\,,
\end{aligned}\label{}
\end{equation}
and for the descendant field,
\begin{equation}
\begin{aligned}
  \tilde{\psi}_{x,z}^{a\mu,e\rho} &= 
    D^{a\mu,e\rho}_{xz} 
  + gf^{abc} \int_{y} D_{xy} \Big(
      \tilde{\psi}^{b\nu,e\rho}_{y,z} 
    	\big(
    		 2\partial_{\nu} \tilde{\mathcal{A}}^{c\mu}_{y} 
    		- \partial_{\mu} \tilde{\mathcal{A}}^{c\nu}_{y} 
    	\big)
    - \tilde{\psi}^{b\mu,e\rho}_{y,z} \partial_{\nu} \tilde{\mathcal{A}}^{c\nu}_{y}
    \Big)
  \\
  &\quad + gf^{abc} \int_{y} D_{xy} \Big(
      \tilde{\mathcal{A}}^{b\nu}_{y} \big(
  		 2\partial^{y}_{\nu} \tilde{\psi}^{c\mu,e\rho}_{y,z}
    	- \partial^{y}_{\mu} \tilde{\psi}_{y,z}^{c\nu,e\rho}
    	\big)
    - \tilde{\mathcal{A}}^{b\mu}_{y} \partial^{y}_{\nu} \tilde{\psi}^{c\nu,e\rho}_{y,z}\Big)
  \\&\quad  
  - g^{2}f^{2}_{ab,cd} \int_{y} D_{xy} 
  	\Big(
    	  \tilde{\psi}^{b\nu,e\rho}_{y,z} \tilde{\mathcal{A}}^{c\mu}_{y} \tilde{\mathcal{A}}^{d\nu}_{y}
  		+ \tilde{\mathcal{A}}^{b\nu}_{y} \tilde{\psi}_{y,z}^{c\mu,e\rho} \tilde{\mathcal{A}}^{d\nu}_{y} 
  		+ \tilde{\mathcal{A}}^{b\nu}_{y} \tilde{\mathcal{A}}_{y}^{c\mu} \tilde{\psi}^{d\nu,e\rho}_{y,z}
  	\Big)\,.
\end{aligned}\label{}
\end{equation}
Further, we consider the descendant equation for the FP ghosts,
\begin{equation}
\begin{aligned}
  \tilde{\gamma}^{a,d}_{x,z} &=  D^{ad}_{xz} - gf^{abc}  \int_{y}D_{xy} 
   \tilde{\mathcal{A}}^{b\mu}_{y} \partial_{\mu} \tilde{\gamma}^{c,d}_{y,z}  \,.
\end{aligned}\label{}
\end{equation}

We now construct the recursion relation for the one-loop level by substituting the perturbiner expansion into the above DS equations. The gluon one-loop off-shell current $\dot{J}^{a\mu}_{\mathcal{P}}$ satisfies
\begin{equation}
\begin{aligned}
  \dot{J}^{a\mu}_{\mathcal{P}}= 
  	  \dot{J}^{a\mu}_{I,\mathcal{P}} 
  	+ \dot{J}^{a\mu}_{II,\mathcal{P}}
  	+ \dot{J}^{a\mu}_{III,\mathcal{P}}
  	+ \dot{J}^{a\mu}_{\text{G},\mathcal{P}} \qquad \text{for}~ |\mathcal{P}|>1\,, 
\end{aligned}\label{}
\end{equation}
where
\begin{equation}
\begin{aligned}
  \dot{J}^{a\mu}_{\text{I},\mathcal{P}} &=
  	-\frac{ig f^{abc}}{k_{\mathcal{P}}^{2}} \sum_{ \mathcal{P} = \mathcal{Q} \cup \mathcal{R}} 
  		\Big(
  			  \tilde{J}_{\mathcal{Q}}^{b \nu} \big(2 k_{\mathcal{R}}^{\nu} \dot{J}_{\mathcal{R}}^{c \mu} -k_{\mathcal{R}}^{\mu} \dot{J}_{\mathcal{R}}^{c \nu}\big)
  			- \tilde{J}_{\mathcal{Q}}^{b\mu} k_{\mathcal{R}}^{\nu} \dot{J}_{\mathcal{R}}^{c \nu}
	  	\\&\qquad\qquad\qquad\qquad\quad
  		+ \dot{J}_{\mathcal{Q}}^{b \nu}\big(2 k_{\mathcal{R}}^{\nu} \tilde{J}_{\mathcal{R}}^{c \mu} -k_{\mathcal{R}}^{\mu} \tilde{J}_{\mathcal{R}}^{c \nu}\big)
  		- \dot{J}_{\mathcal{Q}}^{b \mu} k_{\mathcal{R}}^{\nu} \tilde{J}_{\mathcal{R}}^{c \nu}
  		\Big) 
  	\\&\quad
  		- \frac{g^{2} f^{2}_{a b, c d}}{k_{\mathcal{P}}^{2}} \sum_{\mathcal{P}=\mathcal{Q} \cup \mathcal{R} \cup \mathcal{S}}
  		\Big(
  			  \dot{J}_{\mathcal{Q}}^{b \nu} \tilde{J}_{\mathcal{R}}^{c \nu} \tilde{J}_{\mathcal{S}}^{d \mu}
  			+ \tilde{J}_{\mathcal{Q}}^{b \nu} \dot{J}_{\mathcal{R}}^{c \nu} \tilde{J}_{\mathcal{S}}^{d \mu}
  			+ \tilde{J}_{\mathcal{Q}}^{b \nu} \tilde{J}_{\mathcal{R}}^{c \nu} \dot{J}_{\mathcal{S}}^{d \mu}
  		\Big)\,,
  \\
  \dot{J}^{a\mu}_{\text{II},\mathcal{P}} &= \frac{i g f^{a b c}}{2k_{\mathcal{P}}^{2}} 
  	  \int_{p}
  		\Big(
  			  \big( p^{\nu} + k^{\nu}_{\mathcal{P}} \big) \tilde{\Psi}_{ p | \mathcal{P} }^{ c \nu, b \mu}
  			+ \big( p^{\nu} - 2k^{\nu}_{\mathcal{P}} \big) \tilde{\Psi}_{ p | \mathcal{P} }^{ c \mu, b \nu}
  			- \big( 2p^{\mu} - k^{\mu}_{\mathcal{P}} \big) \tilde{\Psi}_{ p | \mathcal{P} }^{ c \nu, b \nu}
  		\Big)\,,
  \\
  \dot{J}^{a\mu}_{\text{III},\mathcal{P}} &= 
  	- \frac{g^{2} f_{a b, c d}^{2}}{k_{\mathcal{P}}^{2}} \sum_{\mathcal{P}=\mathcal{Q} \cup \mathcal{R}} \int_{p}
  		\Big(
  			  \tilde{J}_{\mathcal{Q}}^{b \nu} \tilde{\Psi}_{p|\mathcal{R}}^{\{c \mu, d \nu\}}
  			+ \tilde{J}_{\mathcal{Q}}^{c \mu} \tilde{\Psi}_{p|\mathcal{R}}^{\{d \nu, b \nu\}}
  			+ \tilde{J}_{\mathcal{Q}}^{d \nu} \tilde{\Psi}_{p|\mathcal{R}}^{\{b \nu, c \mu\}}
  		\Big)\,,
  \\
  \dot{J}^{a\mu}_{\text{G},\mathcal{P}}	 &= \frac{i g f^{abc}}{k_{\mathcal{P}}^{2}} \int_{p}\big(p^{\mu}-k_{\mathcal{P}}^{\mu}\big) \tilde{\Gamma}^{c,b}_{p|\mathcal{P}}\,,
\end{aligned}\label{}
\end{equation}
and the off-shell recursion relation for the descendant currents is
\begin{equation}
  \tilde{\Psi}_{p|\mathcal{P}}^{a\mu,e\rho} = 
  	  \tilde{\Psi}_{\text{I},\mathcal{P}}^{a\mu,e\rho}
  	+ \tilde{\Psi}_{\text{II},\mathcal{P}}^{a\mu,e\rho} \qquad \text{for}~ |\mathcal{P}|>0\,,
\label{}\end{equation}
where
\begin{equation}
\begin{aligned}
  \tilde{\Psi}_{\text{I},\mathcal{P}}^{a\mu,e\rho} &= \frac{igf^{abc}}{\big(p-k_{\mathcal{P}}\big)^{2}} \sum_{\mathcal{P}=\mathcal{Q}\cup\mathcal{R}}
    \bigg[\
    	  \tilde{\Psi}^{b\nu,e\rho}_{p|\mathcal{Q}} \tilde{J}^{c\mu}_{\mathcal{R}}
    		\big( p^{\nu} - k^{\nu}_{\mathcal{Q}} - 2k^{\nu}_{\mathcal{R}} \big)
    	+ \tilde{\Psi}^{b\nu,e\rho}_{p|\mathcal{Q}} \tilde{J}^{c\nu}_{\mathcal{R}}
    		\big( p^{\mu} - k^{\mu}_{\mathcal{Q}} + k^{\mu}_{\mathcal{R}} \big)
    \\&\qquad\qquad\qquad\qquad\qquad
    	- \tilde{\Psi}^{b\mu,e\rho}_{p|\mathcal{Q}} \tilde{J}^{c\nu}_{\mathcal{R}}
    		\big( 2p^{\nu} - 2 k^{\nu}_{\mathcal{Q}} - k^{\nu}_{\mathcal{R}} \big)\
    \bigg]\,,
  \\
  \tilde{\Psi}_{\text{II},\mathcal{P}}^{a\mu,e\rho} &= - \frac{g^{2}f^{2}_{ab,cd}}{\big(p-k_{\mathcal{P}}\big)^{2}}
  \sum_{\mathcal{P}=\mathcal{Q}\cup\mathcal{R}\cup\mathcal{S}}
  	\Big(
    	 \tilde{\Psi}^{b\nu,e\rho}_{p|\mathcal{Q}} \tilde{J}^{c \mu}_{\mathcal{R}} \tilde{J}^{d \nu}_{\mathcal{S}}
    	+ \tilde{\Psi}^{c\mu,e\rho}_{p|\mathcal{Q}} \tilde{J}^{b \nu}_{\mathcal{R}} \tilde{J}^{d \nu}_{\mathcal{S}}
    	+ \tilde{\Psi}^{d\nu,e\rho}_{p|\mathcal{Q}} \tilde{J}^{b \nu}_{\mathcal{R}} \tilde{J}^{c \mu}_{\mathcal{S}}
  	\Big)\,.
\end{aligned}\label{}
\end{equation}
The descendant current for the ghost satisfies
\begin{equation}
\begin{aligned}
  \tilde{\Gamma}^{a,d}_{p|\mathcal{P}} = i g f^{a b c} \frac{1}{(p-k_{\mathcal{P}})^{2}} \sum_{\mathcal{P} = \mathcal{Q} \cup \mathcal{R}} \tilde{J}_{\mathcal{Q}}^{b \mu}\left(p-k_{R}\right)_{\mu} \tilde{\Gamma}_{p | \mathcal{R}}^{c,d} \qquad \text{for}~ |\mathcal{P}|>0\,.
\end{aligned}\label{}
\end{equation}

We now construct the initial condition for the recursion relations. First, we set the initial conditions for the descendant currents, which arise from the zero-mode sector of the descendant equations:
\begin{equation}
  \tilde{\Psi}_{p|\emptyset}^{a\mu,e\rho} = \frac{\delta^{ae}\eta^{\mu\rho}}{p^{2}}\,, \qquad \tilde{\Gamma}_{p|\emptyset}^{a,b} = \frac{\delta^{ab}}{p^{2}}\,.
\label{gluon_first_descendant_current_rank0}\end{equation}
The initial condition for the one-loop gluon current is given by the rank-1 current
\begin{equation}
  \dot{J}^{a\mu}_{i} = \int_{y} D_{xy} j^{\ord{1}}{}^{a\mu}_{y} + \dot{J}^{a\mu}_{\text{II},i} + \dot{J}^{a\mu}_{\text{III},i} = 0 \,,
\label{}\end{equation}
and the initial condition is trivial, $\dot{J}^{a\mu}_{i} = 0$.

\section{Discussion}
We constructed the quantum off-shell recursion relation for computing scattering amplitudes and correlation functions. Firstly, we have derived the quantum perturbiner expansion from the quantum effective action formalism by choosing the external field to reproduce the LSZ reduction formula from the expansion of the connected generating functional $W[j]$. By substituting the perturbiner expansion into the DS equation, we have derived the quantum off-shell recursion relation for $\phi^{4}$-theory and pure Yang--Mills theory. We have introduced the descendant fields and their perturbiner expansions. They are intrinsic loop-level quantities because they carry the off-shell loop momenta explicitly. The number of off-shell legs is increased with the order of descendant fields. We established how to obtain scattering amplitudes from quantum off-shell currents. 

Next, we have derived the one-loop and two-loop recursions for $\phi^{4}$-theory and the one-loop recursion for pure YM theory. Unlike the conventional perturbiner method, the initial condition given by the lowest rank current can be derived from the external source in the DS equation. We explicitly solved the quantum off-shell recursion relation for $\phi^{4}$-theory and checked that the results are the same as the known scattering amplitudes. This work has focused only on solving the scalar off-shell recursion relation. In a follow-up paper, we will compute the loop-level gluon scattering amplitude from the quantum off-shell recursion relation.

We also considered the quantum perturbiner method for correlation functions. The choice of the external source is not unique at all. If we change the form of the source, then the initial condition is also replaced. The new off-shell current leads to another physical quantity. However, the choice of source has no bearing on the off-shell recursion relation itself, which is preserved regardless of the source chosen. We found a form of the external source associated with correlation functions and checked that it reproduces the known one-loop correlation functions. 

We have considered $\phi^{4}$-theory and pure YM theory only. However, this formalism is universal, and we can apply it to any QFT with an action. Especially, it would be interesting to apply it to quantum gravity. In \cite{Cho:2021nim}, the tree-level perturbiner method for gravity was constructed using double field theory (DFT). It would be straightforward to define the quantum perturbiner expansion, and we can construct the quantum recursion relation from the graviton DS equation. This method may provide an alternative tool for computing graviton scattering amplitudes. Further, we can extend this formalism to compute other physical quantities such as in-in correlation functions and correlators for composite fields. These are important for cosmology and the AdS/CFT correspondence.

Finally, applying our quantum perturbiner method to QFTs in curved backgrounds is also interesting, especially black holes, cosmology and AdS spaces, etc. These are important spacetimes for understanding quantum gravity but involve many computations. We expect our recursion relation provides a useful technical tool to analyse QFT in curved backgrounds. We may compute various physical quantities without Feynman diagrams.

\acknowledgments
We thank Kyoungho Cho, Euihun Joung, Minkyoo Kim, Seok Kim, Jeong-Hyuck Park, Jaewon Song for useful discussion and comments. 
This work is supported by appointment to the JRG Program at the APCTP through the Science and
Technology Promotion Fund and Lottery Fund of the Korean Government. It is also supported by
the National Research Foundation of Korea(NRF) grant funded by the Korean government (MSIT)
No.2021R1F1A1060947 and the Korean Local Governments of Gyeongsangbuk-do Province and
Pohang City.

\bibliography{references}

\bibliographystyle{JHEP}

\end{document}